\documentclass[12pt]{article}
\usepackage[dvips]{graphicx}
\usepackage{latexsym}
\usepackage{cite}

\def\Comment#1{}
\def\chpt{$\chi$PT}
\def\rcht{R$\chi$T}
\newcommand{\bean}{\begin{eqnarray*}}
\newcommand{\eean}{\end{eqnarray*}}

\newcommand{\gapproxeq}{\lower
.7ex\hbox{$\;\stackrel{\textstyle >}{\sim}\;$}}
\newcommand{\lapproxeq}{\lower
.7ex\hbox{$\;\stackrel{\textstyle <}{\sim}\;$}}

\newcommand\lsim{\mathrel{\rlap{\lower4pt\hbox{\hskip1pt$\sim$}}
    \raise1pt\hbox{$<$}}}
\newcommand\gsim{\mathrel{\rlap{\lower4pt\hbox{\hskip1pt$\sim$}}
    \raise1pt\hbox{$>$}}}
\newcommand{\ba}{\begin{array}}
\newcommand{\ea}{\end{array}}
\newcommand{\nn}{\nonumber}

\newcommand{\be}{\begin{equation}}
\newcommand{\ee}{\end{equation}}
\newcommand{\bear}{\begin{eqnarray}}
\newcommand{\eear}{\end{eqnarray}}
\newcommand{\tab}{\hspace*{0.5cm}}

\newcommand{\rvac}{\,|0\rangle}

\newcommand{\ket}{\,\rangle}
\newcommand{\bra}{\langle \,}
\newcommand{\eqn}[1]{(\ref{#1})}
\newcommand{\cO}{{\cal O}}
\newcommand{\bel}[1]{\be\label{#1}}
\newcommand{\mL}{\mathcal{L}}
\newcommand{\mA}{\mathcal{A}}

\newcommand{\mF}{\mathcal{F}}

\newcommand{\mI}{\mathcal{I}}

\newcommand{\Frac}[2]{\frac{\displaystyle #1}{\displaystyle #2}}
\newcommand{\Int}{\displaystyle{\int}}
\begin{document}
\thispagestyle{empty}
\begin{titlepage}
\begin{center}
\hfill IFIC/04$-$06\\
\hfill FTUV/04$-$0721 \\
\vspace*{2.75cm}
\begin{Large}
{\bf Quantum Loops in the Resonance Chiral Theory: \\[5pt] The Vector Form Factor}
\\[2.4cm]
\end{Large}
 {\sc I. Rosell}, { \sc J.J. Sanz-Cillero } and { \sc A. Pich }\\[0.8cm]

{\it Departament de F\'\i sica Te\`orica, IFIC, Universitat de Val\`encia -
CSIC\\
 Apt. Correus 22085, E-46071 Val\`encia, Spain }\\[0.5cm]

\today
\vspace*{2cm}
\begin{abstract}
\noindent
We present a calculation of the Vector Form Factor at the next-to-leading order
in the $1/N_C$ expansion, within the framework of Resonance Chiral Theory.
The calculation is performed in the chiral limit, and with two dynamical quark
flavours. The ultraviolet
behaviour of quantum loops involving virtual resonance propagators is analyzed,
together with the kind of counterterms needed in the renormalization procedure.
Using the lowest-order equations of motion, we show that only a few
combinations of local couplings appear in the final result. The low-energy limit
of our calculation reproduces the standard Chiral Perturbation Theory formula,
allowing us to
determine the resonance contribution to the chiral low-energy couplings, at
the next-to-leading order in $1/N_C$, keeping a full control of their
renormalization scale dependence.
\end{abstract}
\end{center}
\vfill
\eject
\end{titlepage}

\pagenumbering{arabic}
\parskip12pt plus 1pt minus 1pt
\topsep0pt plus 1pt
\setcounter{totalnumber}{12}

\section{Introduction}

We have at present an overwhelming experimental and theoretical evidence that
the $SU(3)_C$ gauge theory correctly describes the hadronic world \cite{QCDPich}.
This makes QCD the established theory of the strong interactions. Nevertheless, its
non-perturbative nature at long distances is still challenging our theoretical
capabilities. Since the hadronization procedure is not understood, an
effective field theory description \cite{EFT} in terms of hadronic degrees of freedom is
required in the low-energy regime.

Below the heavy quark thresholds, QCD is properly described considering only the
light quarks, with masses much smaller than the dynamical QCD scale $\Lambda_{QCD}$.
One can then study the massless QCD case and consider the mass term as a perturbation.
With $n_f$ massless quarks, QCD has a chiral $SU(n_f)_R\otimes SU(n_f)_L$ symmetry
which is spontaneously broken to $SU(n_f)_{L+R}$. This generates $(n_f^2-1)$ Goldstone
bosons, which can be identified with the multiplet of light pseudoscalars; their small
masses being proportional to the explicit breaking of chiral symmetry generated by
the quark masses.
The Goldstone nature of the pseudoscalar bosons implies strong constraints on their
interactions, which can be most easily analyzed on the basis of an effective Lagrangian
organized as an expansion in powers of momenta (derivatives) and quark masses over
the chiral symmetry breaking scale $\Lambda_\chi\simeq 4 \pi F_\pi \simeq 1.2$ GeV.
The resulting Goldstone effective field theory,
Chiral Perturbation Theory (\chpt) \cite{WE:79,chpt1loop,chptms}, has achieved a
remarkable success describing the low-energy dynamics of QCD \cite{EC:95,PI:95}.

In the resonance region one must introduce a different effective field theory with
explicit massive fields to describe the degrees of freedom associated with the
mesonic resonances \cite{therole,spin1fields}.
Although chiral symmetry still provides stringent dynamical constraints, the usual
\chpt\ power counting breaks down in the presence of higher energy scales.
Therefore, one needs another expansion parameter to organize the effective Lagrangian.
The limit of an infinite number of quark colours \cite{NC} turns out to be a very
useful tool to understand many features of QCD and provides an alternative power
counting to describe the meson interactions \cite{PI:02}.
Taking $N_C\to\infty$,  with $\alpha_s N_C$ fixed, there exists a systematic
expansion of the $SU(N_C)$ gauge theory in powers of $1/N_C$, which for $N_C=3$ provides
a good quantitative approximation scheme to the hadronic world \cite{MA:98}.
Assuming confinement, the strong dynamics at
$N_C\to\infty$ is given by tree diagrams with infinite sums of hadron exchanges, which
correspond to the tree approximation to some local effective Lagrangian.
Hadronic loops generate corrections suppressed by factors of $1/N_C$.

The short-distance properties of the underlying QCD dynamics impose strong constraints on
the couplings of the hadronic effective theory \cite{spin1fields,PI:02}. The infinite sums
of meson exchanges contributing to any given Green function should obey the right QCD behaviour
$G(t)\sim t^\omega$ at large momenta. This requirement excludes
resonance interactions with large number of derivatives, explaining the phenomenological
success of the usual lowest-order (in derivatives) approximations. Moreover, it implies
stringent correlations among those resonance couplings
associated with the highest powers of momenta.
The usual \chpt\ expansion is of course recovered at very low energies,
when the resonance Green functions are expanded in powers of momenta over
the resonance mass scale.

In spite of its many dynamical simplifications, QCD at $N_C\to\infty$ is
still a very involved theory and some approximations are called for. Usually one
cuts the infinite tower of resonance exchanges to a finite number, taking only
into account those meson states which are relevant at the physical energy scale.
This is meaningful since
the contributions from higher-mass states are suppressed by their corresponding propagators.
However, it introduces back a momentum expansion regulated by inverse powers of
the heavier resonance masses which have been integrated out.
The problem can be formally avoided taking the limit $M_R\to\infty$ for all resonance states
not included in the effective theory. This gives a well-defined approximation with a clear
physical meaning: one is assuming that the QCD short-distance operator product expansion
provides an acceptable description at energies above the last included mesonic state.
The imposed short-distance constraints are nothing else than matching conditions between
the low-energy effective field theory and the underlying QCD dynamics.

The most drastic and simplest scheme is the so-called {\it Single Resonance Approximation},
which only considers the contributions from the lightest meson with any given
quantum numbers \cite{therole,spin1fields,PPdR,GP:00}. The short-distance QCD constraints
determine in this case all hadronic parameters in terms of the pion decay constant $F_\pi$
and the two masses of the vector and scalar multiplets, $M_V$ and $M_S$ \cite{PI:02}.
This gives a very successful description at energies below the scale of the second resonance multiplets.
Since there is an infinite number of Green functions, it is clearly not possible
to satisfy all matching conditions within the single resonance approximation. 
A useful generalization is the
{\it Minimal Hadronic Ansatz}, which keeps the minimum number of resonances
compatible with all known short-distance constraints for the problem at hand \cite{KPdR}.
The resonance contributions to some $\cO(p^6)$ \chpt\ couplings have been already analyzed in this way, by studying an appropriate set of three-point functions \cite{KN:01,RPP:03,CEEPP:04}.

The $N_C\to\infty$ limit of the resonance chiral theory (\rcht) has been investigated in
many works \cite{therole,spin1fields,PI:02,PPdR,GP:00,KPdR,KN:01,RPP:03,CEEPP:04}
and a very successful leading order phenomenology already exists \cite{PI:02}.
However, we are still lacking a systematic procedure to incorporate
next-to-leading contributions in the $1/N_C$ counting.
Up to now, the effort has concentrated in pinning down the most relevant
subleading effects, such as the resonance widths which regulate the corresponding poles
in the meson propagators \cite{GP:97,GPP:00,SP:03}, or the role of final state interactions in
the physical amplitudes \cite{GP:97,GPP:00,SP:03,PP:01,JOP:00,JOP:02,PaP:01,IAM,Palomar}.
These unitarity corrections are generated through
Goldstone loops and therefore are suppressed by $1/N_C$ factors; nevertheless they may
be strongly enhanced by large infrared logarithms.
The combined constraints of analyticity and
unitarity make possible to perform appropriate resummations of chiral logarithms, which
describe the leading $1/N_C$ corrections in the resonance region.

Quantum loops including virtual resonance propagators are a major technical challenge
which has never been properly addressed. A first step in this direction was the study
of resonance loop contributions to the running of the \chpt\ coupling $L_{10}(\mu)$,
performed in Ref.~\cite{CP:01}, which however didn't attempt any analysis of the
induced ultraviolet divergences and their corresponding renormalization.\footnote{
Quantum loops involving massive states have been only analyzed within explicit models
with additional symmetries. For instance, the gauge structure advocated in the so called
``Hidden Local Symmetry'' description of vector resonances \cite{HLS}
implies a much simpler ultraviolet behaviour \cite{HLSloops}.
Chiral loop corrections to some vector resonance parameters have been also studied
\cite{BGT:98,ChR:98} within the context of ``Heavy Vector Meson \chpt'' \cite{JMW:95},
which adopts the $M_R\to\infty$ limit to guarantee a good chiral power counting.
}
Clearly, at the one-loop level the massive states present in \rcht\ generate all kind
of ultraviolet problems which are not yet understood.
A naive chiral power counting indicates that the renormalization procedure will require
higher dimensional counterterms, which presumably could generate a problematic behaviour
at large momenta. Therefore, it will be necessary to perform a careful investigation of
the constraints implied by the short-distance properties of QCD at the
next-to-leading order in $1/N_C$.

A formal renormalization of \rcht\ at the one-loop level appears to be a very involved
task, which requires the prior analysis of several technical ingredients \cite{PPRS:04}.
In order to gain some understanding on the ultraviolet behaviour, it seems worth to
perform first some explicit one-loop calculations of well chosen physical amplitudes.
In the following, we present a detailed investigation of the pion vector form factor
(VFF) at the next-to-leading order (NLO) in the $1/N_C$ expansion.
This observable is defined through the two-Goldstone matrix element of the vector
current:
\be
\bra \pi^+(p_1)\, \pi^-(p_2)\,| \,\Frac{1}{2}\left( \bar{u}\gamma^\mu u -
\bar{d}\gamma^\mu d \right) \rvac \, = \, \mF(q^2) \, (p_1-p_2)^\mu\, ,
\ee
where $q^\mu \equiv (p_1+p_2)^\mu$.
At very low energies, the VFF $\mF(q^2)$ has been studied within the \chpt\ framework
up to $\cO(p^6)$ \cite{MesonFF,op6FF,BT:02}.
\rcht\ and the $1/N_C$ expansion have been also
used to determine $\mF(q^2)$ at the $\rho$ meson peak, including appropriate resummations
of subleading infrared logarithms \cite{GP:97,GPP:00,SP:03,PP:01}.

We will simplify the calculation working in the two flavour theory and taking the
massless quark limit. Therefore, we will assume a chiral $U(2)_L\otimes U(2)_R$ symmetry
group.
The small effects induced by the $U(1)_A$ anomaly will be neglected, because they
are not going to be relevant in our discussion. The isosinglet pseudoscalar can
only appear within loops, and the numerical correction generated by its non-zero mass
could be taken into account in a straightforward way, together with the finite quark mass
effects which we are ignoring.

In the next section we will briefly describe the \rcht\ Lagrangian. We will adopt the
single resonance approximation and will only consider the
minimal set of resonance couplings (linear in the resonance fields) introduced in
ref.~\cite{therole}, supplemented with those counterterms required by the renormalization
procedure.
The renormalization of the relevant one-particle-irreducible Feynman diagrams will be
discussed in section~\ref{sec:renormalization} and the final results of our calculation will
be collected in section~\ref{sec:VFF}. Sections~\ref{sec:LowE} and \ref{sec:LargeE}
analyze the behaviour of the computed vector form factor at low and high energies, respectively.
We will finally summarize our findings in section~\ref{sec:Summary}.
Some technical details have been relegated to the appendices.

\section{The Lagrangian of Resonance Chiral Theory}

We are going to work within a $U(2)_L\otimes U(2)_R$ chiral theory, containing
a multiplet of 4 pseudoscalar Goldstone bosons,
\bel{eq:Goldstones}
\Phi\, =\,\pmatrix{
 {1\over\sqrt 2}\pi^0 + {1\over\sqrt 2}\eta_0 & \pi^+ \cr
\pi^- & - {1\over\sqrt 2}\pi^0 + {1\over\sqrt 2}\eta_0 }\, ,
\ee
parameterized through the unitary matrix \
$u(\Phi) = \exp{\left\{i\,\Phi/(\sqrt{2}\, F)\right\}}$.
The Goldstones couple to massive $U(2)$ multiplets of the type
$V(1^{--})$, $A(1^{++})$, $S(0^{++})$ and $P(0^{-+})$, with a field content
analogous to the one indicated in \eqn{eq:Goldstones}.

Our starting point is the \rcht\ Lagrangian $\mL(u,V,A,S,P)$
introduced in Ref.~\cite{therole}. It contains
the $\cO(p^2)$ \chpt\ Lagrangian~\cite{chpt1loop,chptms},
\bel{eq.L2}
\mL_{2\chi} \, = \, \Frac{F^2}{4} \,\bra u_\mu u^\mu
\, + \, \chi_+
\ket \, ,
\ee
the kinetic resonance Lagrangians,
\bel{eq.LRkin}
\ba{ccl}
\mL_{2Z}(R = V,A) & = &
    - {1\over 2}\, \langle \nabla^\lambda R_{\lambda\mu}
\nabla_\nu R^{\nu\mu} -{1\over 2} \, M^2_R\, R_{\mu\nu} R^{\mu\nu}\rangle
\, ,
\\ \\
\mL_{2Z}
(R = S,P) & = & {1\over 2} \,\langle \nabla^\mu R\,\nabla_\mu R
- M^2_R\, R^2\rangle\, ,
\ea
\ee
and $\cO(p^2)$ interactions linear in the resonance fields:
\bel{eq.LRint}\displaystyle
\ba{ccl}
\mL_{2V}[V(1^{--})] & = &  \Frac{F_V}{ 2\sqrt{2}} \,
     \langle V_{\mu\nu} f_+^{\mu\nu}\rangle\, +\,
    \Frac{i\, G_V}{ 2 \sqrt{2}} \, \langle V_{\mu\nu}
\left[ u^\mu ,  u^\nu \right] \rangle\, ,
\\
\\
\mL_{2A}[A(1^{++})] & = & \Frac{F_A}{ 2\sqrt{2}} \,
    \langle A_{\mu\nu} f_-^{\mu\nu} \rangle\,  ,
\\
\\
\mL_{2S}[S(0^{++})]  & = &  c_d \, \langle S\, u_\mu
u^\mu\rangle\, +\, c_m \, \langle S\, \chi_+ \rangle \,  ,
\\
\\
\mL_{2P}[P(0^{-+})] & = &  i\, d_m \, \langle P\, \chi_-\rangle \,  .
\ea
\ee
The brackets $\bra ... \ket$ denote a trace of the corresponding flavour matrices, and
the different chiral tensors follow the notation defined in Ref.~\cite{therole}.
They include external vector, axial-vector, scalar and pseudoscalar sources
($v^\mu$, $a^\mu$, $s$, $p$) to generate the corresponding Green functions.
Following this reference, we describe the vector and axial-vector resonances
with the antisymmetric field formalism.
In the chiral limit and neglecting external scalar or pseudoscalar sources
$\chi_\pm=0$.

The Lagrangian $\mL(u,V,A,S,P)$ obeys the correct $N_C$ counting rules.
The different fields and the masses and momenta are all of them\ $\cO(1)$,
whereas all couplings ($F$, $F_V$, $G_V$, $F_A$, $c_d$, $c_m$ and $d_m$)
are of\ $\cO\left(\sqrt{N_C}\,\right)$.
In the limit $N_C\to\infty$, one can determine all parameters in terms of $F$,
$M_V$ and $M_S$ \cite{PI:02}. The short-distance QCD behaviour of the vector,
axial-vector \cite{spin1fields} and scalar \cite{JOP:02} form factors, together
with the constraints implied \cite{spin1fields,PI:02,GP:00}
by the superconvergence properties of the\ $vv-aa$ \cite{WE:67} and\ $ss-pp$\
two-point functions at large momenta, imply the relations \cite{PI:02}:
\bel{eq:VASP_coup}
{F_V\over\sqrt{2}} = \sqrt{2}\, G_V = F_A =
2\, c_m = 2\, c_d = 2 \sqrt{2}\, d_m = F \, ,
\ee
\bel{eq:VASP_mass}
M_A = \sqrt{2}\, M_V
\qquad , \qquad
M_P = \sqrt{2}\, M_S \, \left(1 - \delta\right)^{1/2}\, .
\ee
The last identity involves a small correction \
$\delta \approx 3\,\pi\alpha_s F^2/M_S^2 \sim 0.08\,\alpha_s$,
which can be neglected together with the tiny effects from
light quark masses.

\subsection{Subleading Lagrangian}

The one loop calculation of the VFF with the previous Lagrangian generates
ultraviolet divergences which require counterterms with a higher number
of derivatives. We will only include the minimal set of chiral structures
needed to renormalize the VFF calculation. We expect their corresponding
couplings to be subleading in the $1/N_C$ expansion, since they are associated
with quantum loop corrections.

We need to include the following $\cO(p^4)$ and $\cO(p^6)$ Goldstone
interactions:
\bel{eq.L4}
\widetilde{\mL}_{4\chi} \, =\,
\Frac{i\,\widetilde{\ell}_6}{4}\,
\bra f_+^{\mu\nu} \left[ u_\mu , u_\nu \right]  \ket
\, - \, \widetilde{\ell}_{12}\,\bra \nabla^\mu u_\mu \nabla^\nu u_\nu\ket \,  ,
\ee
\bel{eq.L6}
\widetilde{\mL}_{6\chi}
\, = \, i \,\widetilde{c}_{51}\,
\bra \nabla^\rho f_+^{\mu\nu} [h_{\mu\rho},u_\nu] \ket \, + \,
i \,\widetilde{c}_{53}\,
\bra  \nabla_\mu f_+^{\mu\nu} [h_{\nu\rho},u^\rho] \ket \, .
\ee
Chiral invariance forces these terms to have structures contained in the
corresponding \chpt\ Lagrangians \cite{chpt1loop,op6lagrangian}.
We use a tilde to denote the \rcht\ couplings in \eqn{eq.L4} and \eqn{eq.L6},
which are different to the ones with the same names (without tilde) in \chpt.
For instance, the \chpt\ coupling $\ell_6$ ($L_9$ in the three flavour case)
is dominated by a contribution from vector-meson exchange and is of
$\cO(N_C)$, while the corresponding \rcht\ coupling $\widetilde{\ell}_6$
does not contain this contribution and is of $\cO(1)$.

\begin{figure}[th]
\begin{center}
\includegraphics[width=3.5cm,angle=-90,clip]{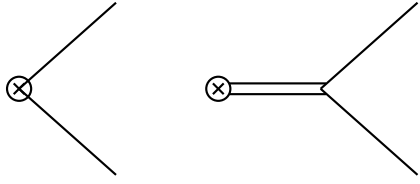}
\caption{\small{Leading order contributions to the VFF.}}
\label{fig.VFFLO}
\end{center}
\end{figure}

The $\cO(p^4)$ term containing the structure $\nabla^\mu u_\mu$
does not contribute to the tree--level calculation of the VFF;
nevertheless, it is needed to renormalize the Goldstone self-energies.
At $\cO(p^6)$, only the combination of couplings
$\widetilde{r}^{\phantom{\, r}}_{V2}\equiv 4 F^2\,(\widetilde{c}_{53}- \widetilde{c}_{51})$
is going to be relevant for the VFF \cite{op6lagrangian}.
Including the Lagrangians \eqn{eq.L4} and \eqn{eq.L6}, the tree level calculation of the
VFF gives the result:
\bel{eq:tree}
\mF(q^2)\, =\, 1 \, +\, {F_V\, G_V\over F^2}\, {q^2\over M_V^2-q^2}\, -\,
\widetilde{\ell}_6\, {q^2\over F^2}\, +\, \widetilde{r}^{\phantom{\, r}}_{V2}\, {q^4\over F^4}\, .
\ee
The requirement that the VFF should vanish at $q^2\to\infty$ implies the following conditions
at leading order in $1/N_C$:
\bel{eq:conditions}
F_V\, G_V\, =\, F^2\qquad , \qquad
\widetilde{\ell}_6\, =\, 0\qquad , \qquad
\widetilde{r}^{\phantom{\, r}}_{V2}\,\equiv\, 4 F^2\,(\widetilde{c}_{53}- \widetilde{c}_{51})\, =\, 0\, .
\ee
Therefore, the couplings $\widetilde{\ell}_6/F^2$ and
$\widetilde{r}^{\phantom{\, r}}_{V2}/F^4$
are of subleading order in the $1/N_C$ expansion, i.e. $\cO(1/N_C)$,
as expected on pure dimensional grounds.

The renormalization of Green functions including resonance fields forces the presence of the
following additional counterterms:
\\[2pt]
\bel{eq.L4VZ}
\ba{rl}
\mL_{4Z} \, =&
\Frac{X_{Z_1}}{2} \,\bra \nabla^2 V^{\mu\nu}
\left\{\nabla_\nu,\nabla^\sigma\right\} V_{\mu\sigma} \ket
\, + \,
\Frac{X_{Z_2}}{4} \,\bra \left\{\nabla_\nu,\nabla_\alpha\right\} V^{\mu\nu}
\left\{\nabla^\sigma,\nabla^\alpha\right\} V_{\mu\sigma} \ket
\\
\\
& \mbox{} +  \,
\Frac{X_{Z_3}}{4} \,\bra \left\{\nabla^\sigma,\nabla^\alpha\right\} V^{\mu\nu}
\left\{\nabla_\nu,\nabla_\alpha\right\} V_{\mu\sigma} \ket\, ,
\ea
\ee
\\[-5pt]
\bel{eq.L4VF}
\mL_{4F}\, = \, X_{F_1} \,\bra V_{\mu\nu} \nabla^2 f_+^{\mu\nu} \ket\,
+\, X_{F_2} \,\bra V_{\mu\nu} \left\{ \nabla^\mu, \nabla_\alpha \right\}
f_+^{\alpha\nu} \ket  \, ,
\ee
\\[-15pt]
\bel{eq.L4VG}
\mL_{4G}\, = \,i\, X_{G_1}\,\bra\left\{ \nabla^\alpha, \nabla_\mu \right\} V^{\mu\nu}
\left[ u_\nu, u_\alpha\right] \ket\, +\,
i\, X_{G_2} \,\bra V^{\mu\nu} \left[ h_{\alpha\mu} , h^\alpha_\nu \right]\ket  \, .
\ee

The quadratic Lagrangian $\mL_{4Z}$ is needed to renormalize the vector self-energy.
Actually, only the sum of couplings \ $X_Z\equiv X_{Z_1}+X_{Z_2}+X_{Z_3}$ \
is relevant for this purpose. The renormalization of the $v$--$V$
two-point function involves the sum of $\mL_{4F}$ couplings
\ $X_F\equiv X_{F_1} + X_{F_2}$.
Finally, the three-point function with one external vector resonance
and two Goldstone legs is renormalized by $\mL_{4G}$ through the combination \
$X_G\equiv   X_{G_2}- \Frac{1}{2} X_{G_1}$.
The dimensions of the couplings are $[X_Z]=E^{-2}$ and $[X_{F}]=[X_{G}]= E^{-1}$.

At the NLO in $1/N_C$, these counterterm Lagrangians only contribute
through tree-level diagrams. One can then use the leading order equation of motion (EOM)
\bel{eq.EOM}
\ba{rl}
\nabla^\mu \nabla_\rho V^{\rho\nu} - \nabla^\nu \nabla_\rho V^{\rho\mu}
\, =& \, -  M_V^2\, V^{\mu\nu}\, -\, \Frac{F_V}{\sqrt{2}}\, f_+^{\mu\nu}\,
-\, \Frac{i G_V}{\sqrt{2}}\,\left[u^\mu,u^\nu\right]
\ea
\ee
to reduce the number of relevant operators.
The Lagrangians \eqn{eq.L4VZ}, \eqn{eq.L4VF}  and \eqn{eq.L4VG}
take then the equivalent forms:
\\[2pt]
\bel{eq.L4VZEOM}
\ba{rl}
\mL_{4Z}^{\mathrm{EOM}} \, =&
\Frac{X_Z M_V^4}{2} \,\bra V^{\mu\nu} V_{\mu\nu} \ket
\, +\,
\Frac{X_Z M_V^2 F_V}{\sqrt{2}} \,\bra V_{\mu\nu} f_+^{\mu\nu} \ket
\, + \,  \Frac{i\, X_Z M_V^2 G_V}{\sqrt{2}} \,\bra V_{\mu\nu}
\left[ u^\mu , u^\nu \right] \ket
\\
\\
&\mbox{} + \,
\Frac{ i\, X_Z F_V G_V}{2} \,\bra f_+^{\mu\nu} \left[ u_\mu , u_\nu \right]
\ket\, + \,\cdots\, ,
\ea
\ee
\\[-7pt]
\bel{eq.L4VFEOM}
\mL_{4F}^{\mathrm{EOM}} \, = \,
- X_{F} M_V^2\,  \bra V_{\mu\nu} f_+^{\mu\nu} \ket\,
-\, \Frac{i\, X_{F} G_V}{\sqrt{2}} \,\bra f^{\mu\nu}_+
\left[ u_\mu , u_\nu \right]  \ket\, + \,\cdots \, ,
\ee
\\[-7pt]
\bel{eq.L4VGEOM}
\mL_{4G}^{\mathrm{EOM}} \, = \, - 2  i\, X_{G} M_V^2\,
\bra V^{\alpha\nu} \left[ u_\alpha, u_\nu \right] \ket\,
-\, i\, \sqrt{2}\,  X_{G} F_V\,
\bra f_+^{\mu\nu} \left[ u_\mu , u_\nu \right] \ket\, + \,\cdots\, ,
\ee
\\[2pt]
where the dots denote other terms which are not relevant
for the VFF calculation at this order.
The derivatives acting on the vector resonance fields have been traded
by the heavy mass scale $M_V$ and/or derivatives acting on the
Goldstone fields, giving rise to the usual tensor structures of the
\chpt\ Lagrangian. Therefore, the effect of the
counterterm Lagrangians $\mL_{4Z}$, $\mL_{4F}$ and $\mL_{4G}$
is just equivalent to the following shift in the couplings
at the next-to-leading order in $1/N_C$:
\bel{eq.effcouplings}
\ba{rl}
\widetilde{\ell}_6^{\,\mathrm{eff}} \, =& \widetilde{\ell}_6 \, + \,
2\, X_Z F_V G_V  \, - \, 2 \sqrt{2}\, X_F G_V
\, - \, 4 \sqrt{2}\, X_G F_V  \, ,
\\
\\
F_V^{\,\mathrm{eff}} \, =& F_V \, + \, 2\, X_Z M_V^2 F_V
\, - \, 2 \sqrt{2}\, X_{F} M_V^2 \, ,
\\
\\
G_V^{\,\mathrm{eff}} \, =& G_V \, + \, 2\, X_Z M_V^2 G_V
\, - \, 4 \sqrt{2}\, X_{G} M_V^2 \, ,
\\
\\
(M_V^2)^{\mathrm{eff}} \, =& M_V^2 \, + \, 2\, X_Z M_V^4\, ,
\\
\\
\widetilde{r}^{\,\mathrm{eff}}_{V2}\, =&  \widetilde{r}^{\phantom{\, r}}_{V2}\, .
\ea
\ee
Thus, since $\widetilde{\ell}_6^{\,\mathrm{eff}} \sim\widetilde{\ell}_6
\sim (M_V^2)^{\mathrm{eff}}\sim M_V^2\sim\cO(1)$ and
$F_V^{\,\mathrm{eff}}\sim F_V\sim G_V^{\,\mathrm{eff}}\sim G_V\sim\cO(\sqrt{N_C})$,
a consistent $1/N_C$ counting requires that
$X_G$ and $X_F$ are of $\cO(1/\sqrt{N_C})$ and $X_Z$ of $\cO(1/N_C)$.

\section{Renormalization of Quantum Loops}
\label{sec:renormalization}
\tab
The renormalization procedure follows very systematic and precise steps
in any well defined field theory.
First of all, the two-point Green functions must
be renormalized. Later the three-point Green functions and so on. For the VFF up
to NLO in $1/N_C$ only the two- and three-point Green
functions will contribute. The corresponding renormalizations for the one
particle irreducible diagrams (1PI) at one loop are given in the next
subsections.

We will adopt the $\overline{MS}-1$ scheme, usually employed in
\chpt\ calculations, where one subtracts the divergent constant
\be
\lambda_\infty\, =\,\frac{2\, \mu^{d-4}}{d-4}+\gamma_E-\ln{4\pi}-1\, .
\ee
However, we will impose the on-shell condition to renormalize the pion
self-energy. This simplifies the calculation of physical amplitudes with
external pions.
Since we work in the massless quark limit, the Goldstone tadpoles will not give
any contribution.
The precise definition of the relevant Feynman integrals with one, two
and three propagators and some useful antisymmetric formalism technology
are relegated to appendices A and B.

\subsection{Pion self-energy}
\begin{figure}[h]\centering   
\includegraphics[height=12cm,angle=-90,clip]{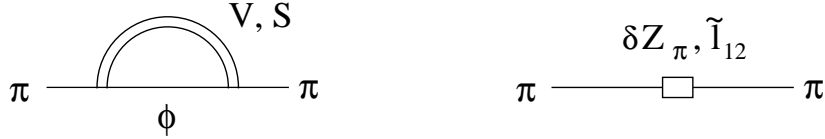}
\caption{\small One-loop diagrams and local contributions to the pion self-energy.}
\label{fig.piself}
\end{figure}

The diagrams contributing to the pion propagator are shown in Fig.~\ref{fig.piself}.
The kinetic Lagrangians $\mL_{2Z}(R)$ in Eq.~\eqn{eq.LRkin} generate additional
tadpole topologies with one resonance propagator, but they are identically zero
even with massive pions.
The divergences of $\cO(p^2)$ are reabsorbed through the
wave-function renormalization \
\mbox{$\pi^{(B)}=(1+\delta Z_\pi)^\frac12\,\pi^{(r)}$},
being $\pi^{(B)}$ and $\pi^{(r)}$
the bare and renormalized pion fields respectively.
In the on-shell scheme,
\be
\delta Z_\pi \, = \, -\Frac{2\, G_V^2}{F^2}\,
\Frac{3\, M_V^2}{16\pi^2 F^2}\,
\left\{ \lambda_\infty + \ln{\Frac{M_V^2}{\mu^2}} + \Frac{1}{6} \right\}
\, + \, \Frac{4\, c_d^2}{F^2}\,\Frac{M_S^2}{16\pi^2 F^2}\,
\left\{\lambda_\infty + \ln{\Frac{M_S^2}{\mu^2}} - \Frac{1}{2}\right\}\, .
\ee
There are also divergences of $\cO(p^4)$ which renormalize
one of the couplings in $\widetilde{\mL}_{4\chi}$:
\be
\widetilde{\ell}_{12}\,\equiv\, \widetilde{\ell}_{12}^r(\mu) +
\delta \widetilde{\ell}_{12}(\mu)
\qquad\quad ; \qquad\quad
\delta \widetilde{\ell}_{12}(\mu) \, = \,
-\Frac{G_V^2 + 2\, c_d^2}{F^2}\;
\Frac{\lambda_\infty}{32\pi^2}\; .
\ee
The renormalized pion self-energy takes the form
\bear
-i\,\Sigma_\pi^r(p^2)& = &
 -i\, {p^4\over 16\pi^2 F^2}\,\left\{
64\pi^2 \,\widetilde{\ell}_{12}^r(\mu) \,
+ \, \Frac{2\, G_V^2}{F^2}
\left[\ln{\Frac{M_V^2}{\mu^2}} + \phi\left(\Frac{p^2}{M_V^2}\right)\right]
\right.\nn\\[10pt] &&\left.\hskip 2cm\mbox{}
+ \, \Frac{4c_d^2}{F^2}\left[\ln{\Frac{M_S^2}{\mu^2}}
+ \phi\left(\Frac{p^2}{M_S^2}\right)\right]\right\} \, ,
\eear
where the function
%
\be
\phi(x) = \left(1-{1\over x}\right)^2 \left[ \left(1-{1\over x}\right)
\ln{(1-x)} -1 + {x\over 2}\right]
\; = \; - (1-x)^2 \,\sum_{n=0}^\infty\, {x^n\over (n+2)(n+3)}
\ee
contains finite and scale-independent contributions.

\subsection{Rho self-energy}
\begin{figure}[h]  
\begin{center}
\includegraphics[height=11cm,angle=-90,clip]{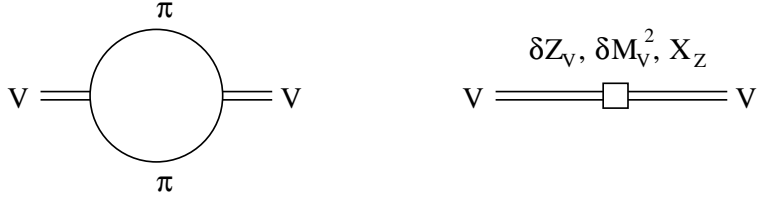}
\caption{\small{Rho self-energy.}}
\label{fig.rhoself}
\end{center}
\end{figure}

The one loop $\rho$ self-energy contains only an $\cO(p^4)$ divergence, which
renormalizes the coupling $X_{Z}$ of the NLO resonance Lagrangian:
\bel{eq:XZ_ren}
X_{Z}\,\equiv\, X_Z^r(\mu) + \delta X_{Z}(\mu)
\qquad\quad ;  \qquad\quad
\delta X_{Z}(\mu)\, =\, -\Frac{2\, G_V^2}{F^2}\,
\Frac{\lambda_\infty}{192 \pi^2 F^2}\; .
\ee
Thus, the vector mass and wave-function are not renormalized:
\be
\delta M_V^2 \, =\, 0
\qquad\quad ;  \qquad\quad
\delta Z_V \, = \, 0 \, .
\ee
The renormalized $\rho$ self-energy then becomes:
\bel{eq.rhoSE}
-i\,\Sigma_V^r(q)^{\mu\nu,\rho\sigma} \, =\,
-{i\over 2}\,\mA^{\mu\nu,\rho\sigma}(q)\;\Sigma_V^r(q^2)\, ,
\ee
where the antisymmetric tensor structure $\mA^{\mu\nu,\rho\sigma}(q)$
is defined in appendix B and
\be\label{eq:VM_SE}
\Sigma_V^{\,r}(q^2)\, =\,
- q^4\,\left\{ 2\, X^r_Z(\mu)
\, - \,\Frac{2\, G_V^2}{F^2}\,\Frac{1}{F^2}\,\left[
{1\over 6}\,\hat{B}_0(q^2/\mu^2) + \Frac{1}{144\pi^2} \right]\right\} \, .
\ee

\subsection{$\langle v^\mu\, V^{\rho\sigma}\rangle$ 1PI vertex}
\begin{figure}[h]
\begin{center}
\includegraphics[height=10cm,angle=-90,clip]{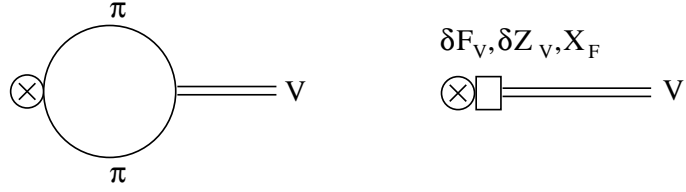}
\caption{\small{Diagrams contributing to the
$\langle v^\mu\, V^{\rho\sigma}\rangle$ Green function
at NLO in $1/N_C$.}}
\label{fig.FV}
\end{center}
\end{figure}

The 1PI amputated diagrams (at NLO) connecting an external vector quark current
to an outgoing vector resonance are shown in Fig.~\ref{fig.FV}.
The one-loop contribution brings an $\cO(p^4)$ divergence which gets reabsorbed
through the following renormalization of the coupling $X_F$:
\be
X_{F}\,\equiv\, X_F^r(\mu) + \delta X_F(\mu)
\qquad\quad ;  \qquad\quad
\delta X_F(\mu)\, =\, -\Frac{\sqrt{2}\, G_V}{F}\,
\Frac{\lambda_\infty}{192 \pi^2 F}\; .
\ee
Since there are no divergences of $\cO(p^2)$, the lowest-order coupling
$F_V$ remains unchanged:
\be
\delta F_V \, =\, 0 \, .
\ee
The renormalized vertex function takes the form
\be
i \,\Phi (q)^{\mu,\rho\sigma} \, = \,
- i \, \mI^{\rho\sigma}_{\alpha\beta}\;
q^{\alpha} g^{\mu\beta} \, \left\{ F_V
\, - 2 \sqrt{2}\, X_F^r(\mu) \, q^2 \, + \, \Frac{2 G_V}{F^2} \, q^2\,
\left[\Frac{1}{6}\,\hat{B}_0(q^2/\mu^2) + \Frac{1}{144\pi^2} \right]\right\} \, ,
\ee
where the first term is the leading order contribution.
The massless two-point function $\hat{B}_0(q^2/\mu^2)$ is defined in appendix A and
the antisymmetric tensor structure $\mI^{\rho\sigma}_{\alpha\beta}$
in appendix~B.

\subsection{$\langle V_{\mu\nu}\pi\pi\rangle$ 1PI vertex}
\begin{figure}[th]
\begin{center}
\includegraphics[height=15cm,angle=-90,clip]{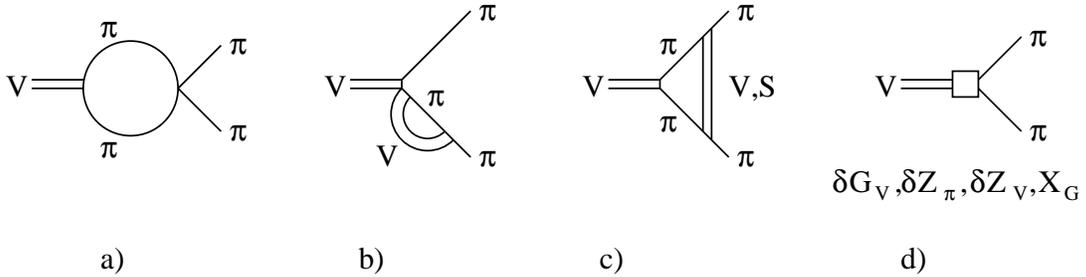}
\caption{\small{NLO diagrams contributing to the three-point Green function
$V^{\mu\nu}\to \pi\pi$.}}
\label{fig.GV}
\end{center}
\end{figure}

The 1PI amputated diagrams connecting
a vector resonance with two outgoing pions
at NLO in $1/N_C$ are shown in Fig.~\ref{fig.GV}.
The loop diagrams generate $\cO(p^2)$ and $\cO(p^4)$ divergences,
which renormalize the couplings $G_V$ and $X_{G}$, respectively:
\be
G_V\,\equiv\, G_V^r(\mu) + \delta G_V(\mu)
\qquad ;  \qquad
\delta G_V(\mu) \,=\,
 G_V  \left[
3\, M_V^2\,\left( \Frac{2 G_V^2}{F^2}-\Frac{1}{2}\right)
- M_S^2\, \Frac{4 c_d^2}{F^2} \right] \Frac{\lambda_\infty}{16\pi^2 F^2} \; .
\ee
\be
X_G\,\equiv\, X_G^r(\mu) + \delta X_G(\mu)
\qquad ;  \qquad
\delta X_G(\mu)\, =\,
\Frac{\sqrt{2}\, G_V}{F}  \, \left[ \Frac{2 G_V^2}{F^2}
+ \Frac{4 c_d^2}{F^2} - 2 \right]
\Frac{ \lambda_\infty}{1536\pi^2 F} \; .
\ee
The wave-function renormalization of the external vector and pion legs amounts
to a global factor
$\left(\delta Z_\pi+\frac{1}{2} \,\delta Z_V\right)$
multiplying the lowest-order contribution
($\delta Z_V=0$ at NLO).
Taking this into account, one finally gets the
finite vertex function
\be
i \,\Gamma_{\mu\nu}^{\,r}(p_1,p_2)\, =\,
\mathcal{I}^{\alpha\beta}_{\mu\nu}\; q_\alpha (p_1-p_2)_\beta\,
\Frac{1}{F^2}\,\left\{
G_V^r(\mu) \,\left[ 1 -\Delta\Gamma (q^2,\mu^2)\right]
\, -\, 4\sqrt{2}\, X_G^r(\mu)\, q^2\right\}\, ,
\ee
where
\bear\label{eq:DGamma}
\Delta\Gamma (q^2,\mu^2) &\!\! = &\!\! {1\over F^2}\;\left\{\;
\hat{B}_0(q^2/\mu^2)\;\left[
\Frac{2 G_V^2}{F^2} \left(\Frac{M_V^4}{q^2}+2 M_V^2 + \Frac{q^2}{12} \right)
\, +\, \Frac{4 c_d^2}{F^2}\left(\Frac{M_S^4}{q^2}+\Frac{q^2}{12}\right)
-\Frac{q^2}{6}\right]
\right.\nn\\[10pt] &&\mbox{}
+\, \Frac{M_V^2}{16\pi^2}\,\ln{\Frac{M_V^2}{\mu^2}}\,
\left[ \Frac{2G_V^2}{F^2}\left( \Frac{M_V^2}{q^2}+5\right)
-\Frac{3}{2} \right]
\; +\; \Frac{M_S^2}{16\pi^2}\,\ln{\Frac{M_S^2}{\mu^2}}\;\,
\Frac{4c_d^2}{F^2}\left(\Frac{M_S^2}{q^2}-1\right)
\nn\\[10pt] &&\mbox{}
+\, \Frac{M_V^2}{64\pi^2} \left[ 3\,\Frac{2G_V^2}{F^2}-1\right]
\; +\;\Frac{3M_S^2}{64\pi^2}\,\Frac{4c_d^2}{F^2}
\; +\;\Frac{q^2}{288\pi^2}\left[\Frac{2G_V^2}{F^2}+\Frac{4c_d^2}{F^2}-2\right]
\nn\\[10pt] &&\mbox{}
+\,\Frac{2 G_V^2}{F^2}\;
C_0(q^2,0,0,M_V^2) \;\left[ \Frac{M_V^6}{q^2}+\Frac{5M_V^4}{2}+q^2 M_V^2\right]
\nn\\[10pt] &&\mbox{}\left.
+\, \Frac{4c_d^2}{F^2}\;
C_0(q^2,0,0,M_S^2)\;\left[\Frac{M_S^6}{q^2}+\Frac{M_S^4}{2}\right]\;
\right\} \, .
\eear
The three-propagator integral $C_0(q^2,M_a^2,M_b^2,M_c^2)$ is defined in appendix A.

\subsection{$\langle v_\mu\pi\pi\rangle$ 1PI vertex}

\begin{figure}[t]
\begin{center}
\includegraphics[height=16cm,angle=-90,clip]{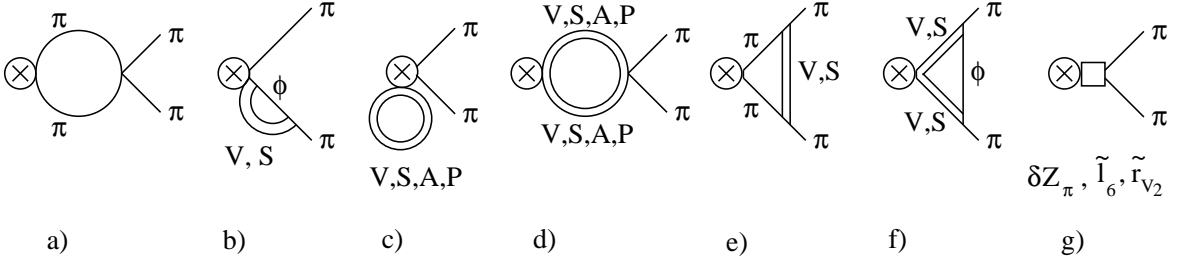}
\caption{\small{1PI diagrams
connecting an external vector current and two outgoing pions.}}
\label{fig.VFFl6}
\end{center}
\end{figure}

The divergences generated by the 1PI loop diagrams shown in
Fig.~\ref{fig.VFFl6} get reabsorbed through the renormalization
of the pion wave function $\delta Z_\pi$ and the
$\cO(p^4)$ and $\cO(p^6)$  couplings $\widetilde{\ell}_6$
and $\widetilde{r}^{\phantom{\, r}}_{V2}$:
\bear\label{eq:l6_run}
\widetilde{\ell}_6 \,\equiv\,\widetilde{\ell}^r_6(\mu)
+ \delta \widetilde{\ell}_6(\mu)
\qquad &;&\qquad
\delta \widetilde{\ell}_6(\mu) \, =\, \left\{
3-2\,\Frac{2G_V^2}{F^2}+\Frac{4c_d^2}{F^2}
\right\} \,
\Frac{\lambda_\infty}{96\pi^2} \; ,
\\
\widetilde{r}^{\phantom{\, r}}_{V2}\,\equiv\,\widetilde{r}^{\, r}_{V2}(\mu) +
\delta \widetilde{r}^{\phantom{\, r}}_{V2}(\mu)
\qquad &;&\qquad
\delta \widetilde{r}^{\phantom{\, r}}_{V2}(\mu) \, =\,
\Frac{F^2\lambda_\infty}{96 \pi^2}\, \left\{
\Frac{1}{M_V^2} + \Frac{1}{M_A^2}\right\} \, .
\label{eq:f1_run}\eear
The resulting finite correction to the lowest-order pion form factor,
\be\label{eq:DF_1PI}
\Delta\mF(q^2)_{\mathrm{1PI}} \, = \,\Delta\mF^{\,\mathrm{ct}}_\chi\, +\,
\Delta\mF^{\mL_2}_\chi\, +\, \Delta\mF^{V}_\chi\, +\, \Delta\mF^A_\chi
\, +\,\Delta\mF^S_\chi\, + \,\Delta\mF^{P}_\chi\, ,
\ee
contains contributions from tree-level counterterms,
\bear\label{eq.finitol6Zpi}
\Delta\mF^{\,\mathrm{ct}}_\chi & =&
-\,\Frac{2 G_V^2}{F^2}\,\Frac{M_V^2}{16\pi^2 F^2}
\left\{ 3\,\ln{\Frac{M_V^2}{\mu^2}} + \Frac{1}{2} \right\}
 \, + \,  \Frac{4c_d^2}{F^2}\,\Frac{M_S^2}{16\pi^2 F^2}\left\{
\ln{\Frac{M_S^2}{\mu^2}} - \Frac{1}{2}\right\}
\nn\\[10pt] &&\mbox{}
- \, \widetilde{\ell}^r_6(\mu)\,\Frac{q^2}{F^2}
\, +\, \widetilde{r}^{\, r}_{V2}(\mu)\,\Frac{q^4}{F^4} \, ,
\eear
and loop diagrams with internal Goldstone bosons
(first diagram in Fig.~\ref{fig.VFFl6}),
\bel{eq.finitol6L2}
\Delta\mF^{\mL_2}_\chi \, =\,\Frac{q^2}{F^2}
\left\{\Frac{1}{6}\,\hat{B}_0(q^2/\mu^2) + \Frac{1}{144\pi^2}\right\} \, ,
\ee
and vector,
\bear\label{eq.finitol6V}
\Delta\mF^{V}_\chi & =&
\Frac{2G_V^2}{F^2} \,\Frac{1}{F^2} \left\{
-\, C_0(q^2,0,0,M_V^2) \,
\left[\Frac{M_V^6}{q^2} + \Frac{5 M_V^4}{2} + q^2
M_V^2\right]
\right.\nn\\[10pt] &&\left.\hskip 1.5cm\mbox{}
+\, C_0(q^2,M_V^2,M_V^2,0) \,
\left[\Frac{M_V^6}{q^2} + \Frac{M_V^4}{2} \right]
\, -\, \hat{B}_0(q^2/\mu^2) \,
\left[\Frac{M_V^4}{q^2}+2M_V^2+\Frac{q^2}{12}
\right]\right\}
\nn\\[10pt] &&\mbox{} -\,
\Frac{\overline{B}_0(q^2,M_V^2)}{F^2} \, \left[
\left(2 M_V^2+\Frac{q^2}{6}-\Frac{q^4}{6 M_V^2}\right)\,
+\,
\Frac{2G_V^2}{F^2}\left(\Frac{M_V^4}{q^2}
+ \Frac{2 M_V^2}{3} - \Frac{5 q^2}{12}\right) \right]
\\[10pt] &&\mbox{} +\,
\Frac{M_V^2}{16\pi^2 F^2}\,\ln{\Frac{M_V^2}{\mu^2}}\,
\left[  \left(\Frac{q^2}{2 M_V^2}-\Frac{q^4}{6 M_V^4} \right)\,
-\, \Frac{2 G_V^2}{F^2} \left(  \Frac{M_V^2}{q^2}-1
+ \Frac{5q^2}{12M_V^2} \right) \right]
\nn\\[10pt] &&\mbox{} +\,
\Frac{M_V^2}{16\pi^2 F^2} \left[
\left(\Frac{q^2}{2 M_V^2}\, -\,\Frac{2 q^4}{9 M_V^4}\right)
\, +\,\Frac{2G_V^2}{F^2}\left(
\Frac{M_V^2}{q^2}\, +\, 1\, -\,\Frac{19 q^2}{36 M_V^2}\right)
\right] \, ,
\nn\eear
axial-vector,
\bear\label{eq.finitol6A}
\Delta\mF^A_\chi & =&\mbox{}-\,
\Frac{\overline{B}_0(q^2,M_A^2)}{F^2}\left[ 2M_A^2+\Frac{q^2}{6}-\Frac{q^4}{6M_A^2}
\right]\, + \, \Frac{M_A^2}{16\pi^2F^2} \ln{\Frac{M_A^2}{\mu^2}}\,
\left[\Frac{q^2}{2M_A^2}-\Frac{q^4}{6M_A^4} \right]
\nn\\ &&\mbox{}
 +\,\Frac{q^2}{32\pi^2F^2}\, -\,\Frac{q^4}{72\pi^2 F^2 M_A^2} \, ,
\eear
scalar,
\bear\label{eq.finitol6S}
\Delta\mF^S_\chi & =&\mbox{}
\Frac{4c_d^2}{F^2} \,\Frac{1}{F^2} \left\{
-\, C_0(q^2,0,0,M_S^2)\,
\left[\Frac{M_S^6}{q^2}+\Frac{M_S^4}{2} \right]
\, +\, C_0(q^2,M_S^2,M_S^2,0)\,
\left[\Frac{M_S^6}{q^2}-\Frac{M_S^4}{2}\right]
\right.\nn\\[10pt] &&\hskip 1.5cm\left.
\, -\,\hat{B}_0(q^2/\mu^2) \,
\left[\Frac{M_S^4}{q^2}+\Frac{q^2}{12} \right]
\, +\,\Frac{M_S^4}{16\pi^2 q^2}\right\}
\quad\; -\quad\;
\Frac{q^2}{288\pi^2F^2}\left[ 1+\Frac{1}{2}\,\Frac{4c_d^2}{F^2}\right]
\nn\\[10pt] &&\mbox{}
-\, \Frac{\overline{B}_0(q^2,M_S^2)}{F^2}\,
\left[\left(\Frac{2M_S^2}{3} - \Frac{q^2}{6}\right)\, +\,\Frac{4 c_d^2}{F^2}\,
\left(\Frac{M_S^4}{q^2}-\Frac{M_S^2}{3}+\Frac{q^2}{12}\right)\right]
\\[10pt] &&\mbox{}
-\, \Frac{M_S^2}{16\pi^2F^2}\,\ln{\Frac{M_S^2}{\mu^2}}\,
 \left[ \Frac{4c_d^2}{F^2}\,
\left(1+\Frac{M_S^2}{q^2} -\Frac{q^2}{12M_S^2}\right)\, +\,
\Frac{q^2}{6M_S^2} \right]\, ,
\nn\eear
and pseudoscalar resonances,
\bel{eq.finitol6P}
\Delta\mF^{P}_\chi \, =\,
\Frac{\overline{B}_0(q^2,M_P^2)}{F^2} \left[ -\Frac{2M_P^2}{3}+\Frac{q^2}{6} \right]
\, - \, \Frac{q^2}{96\pi^2F^2} \left[ \ln{\Frac{M_P^2}{\mu^2}}+\Frac{1}{3}\right] \, .
\ee

\section{Vector Form Factor}
\label{sec:VFF}
\begin{figure}[h]\centering
\centerline{
\begin{minipage}[c]{4.5cm}\centering  
\includegraphics[height=4.5cm,angle=-90,clip]{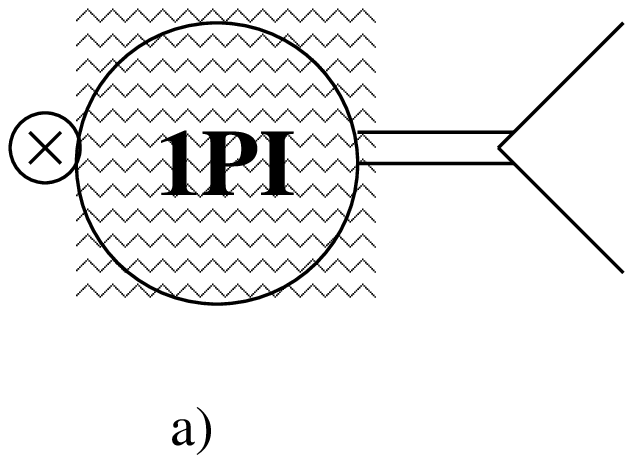}
\end{minipage}
\hskip 1cm  
\begin{minipage}[c]{4.5cm}\centering  
\includegraphics[height=4.5cm,angle=-90,clip]{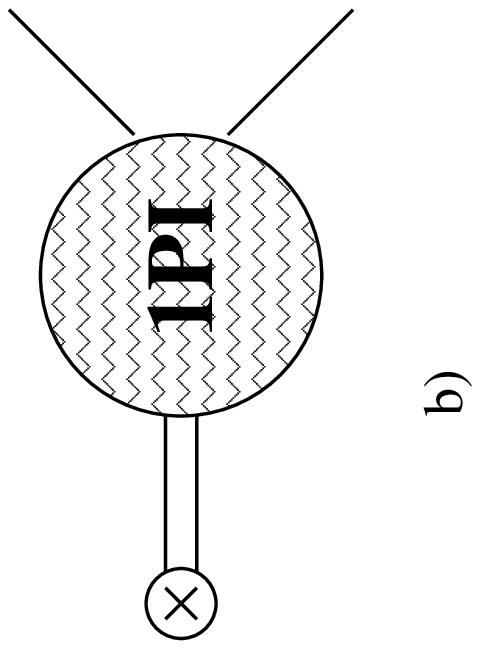}
\end{minipage}
\hskip 1cm  
\begin{minipage}[c]{4.5cm}\centering  
\includegraphics[height=4.5cm,angle=-90,clip]{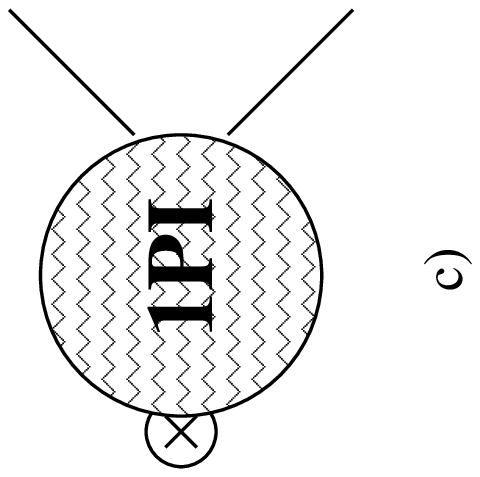}
\end{minipage}}
\caption{\small{Basic topologies contributing to the Vector Form Factor at NLO.}}
\label{fig:topologies}
\end{figure}

The basic topologies contributing to the VFF are shown in
Fig.~\ref{fig:topologies}, in terms of the NLO 1PI diagrams computed
in the previous section. The internal $\rho$ line denotes the dressed
vector propagator, including the self-energy correction~\eqn{eq:VM_SE}
which regulates the $\rho$ pole. Taking this self-energy into account,
the LO contribution takes the form:
\bel{eq:VFF_LO}
\mF(q^2)_{\mathrm{LO}}\, =\, 1 \, +\, \Frac{F_V\, G_V^{\,r}(\mu)}{F^2}\;
\Frac{q^2}{M_V^2-q^2 - \Sigma^{\,r}_V(q^2)}\; .
\ee

The topology in Fig.~\ref{fig:topologies}.a generates the following NLO
correction:
\bel{eq:DFF}
\Delta\mF(q^2)_F \, = \,
\Frac{q^2}{M_V^2-q^2-\Sigma^{\,r}_V(q^2)} \;\Frac{q^2}{F^2}\;\left\{
\Frac{2 G_V^2}{F^2}\,
\left[\Frac{1}{6}\,\hat{B}_0(q^2/\mu^2) + \Frac{1}{144\pi^2}\right]
-2\, F\, {\sqrt{2} G_V\over F}\, X_F^r(\mu)\right\}\, .
\ee
The second topology (Fig.~\ref{fig:topologies}.b) brings the contribution:
\bel{eq:DFG}
\Delta\mF(q^2)_G \, = \, -\,
\Frac{q^2}{M_V^2-q^2-\Sigma^{\,r}_V(q^2)} \;\Frac{F_V}{\sqrt{2} F}
\;\left\{
\Frac{\sqrt{2} G_V}{F}\;\Delta\Gamma(q^2,\mu^2) \, + \,
\Frac{8\, X_G^r(\mu)}{F}\, q^2\right\}\, ,
\ee
where $\Delta\Gamma(q^2,\mu^2)$ is given in Eq.~\eqn{eq:DGamma}.
Finally, Fig.~\ref{fig:topologies}.c denotes the 1PI correction
$\Delta\mF(q^2)_{\mathrm{1PI}}$  in Eq.~\eqn{eq:DF_1PI}.
Adding all contributions together, one gets the VFF at NLO:
\be
\mF(q^2) \, = \, \mF(q^2)_{\mathrm{LO}} \, + \, \Delta\mF(q^2)_{F} \, + \,
\Delta\mF(q^2)_G \, + \,\Delta\mF(q^2)_{\mathrm{1PI}}\, .
\ee

Using the large--$N_C$ relations \eqn{eq:VASP_coup}
in the NLO terms, the result can be written in the form:
\bel{eq:VFF_result}
\mF(q^2) \; = \; A(q^2)\;\, {M_V^2\over M_V^2-q^2-\Sigma^{\,r}_V(q^2)}\; +\; B(q^2)\, ,
\ee
where
\bear
A(q^2) & = & 1 + \hat{\delta}_V + 2\, M_V^2\, \hat{X} - \Delta\tilde{\Gamma}(q^2)\, ,
\nn\\[10pt]
B(q^2) & = & \mathcal{G}(q^2) - \hat{\delta}_V - 2\, (M_V^2+q^2)\, \hat{X}\, .
\eear
The constants
\bear
\hat{\delta}_V&\equiv & {F_V\, G_V^r(\,\mu)\over F^2} - 1 - \Delta\Gamma (0,\mu^2)\, ,
\nn\\[10pt]
\hat{X}&\equiv & X_Z^r(\mu) -{1\over F}\,\left[ X_F^r(\mu) + 4\, X_G^r(\mu)\right]
\eear
and the functions $\Sigma^{\,r}_V(q^2)$,
\be
\Delta\tilde{\Gamma}(q^2)\;\equiv\;
\Delta\Gamma (q^2,\mu^2)\, -\,\Delta\Gamma (0,\mu^2)\, ,
\ee
and
\be
\mathcal{G}(q^2)\;\equiv\; \Delta\mF(q^2)_{\mathrm{1PI}}\, +\,
\Delta\tilde{\Gamma}(q^2)
\;\equiv\; G(q^2,\mu^2) \, -\,\Delta\Gamma (0,\mu^2)
\ee
are independent of the renormalization scale $\mu$.
The subleading \rcht\ couplings $X_F^r(\mu)$ and $X_G^r(\mu)$ only appear
through the constant $\hat{X}$, while $X_Z^r(\mu)$ is also present
in the function $\Sigma^{\,r}_V(q^2)$.
At $q^2=0$, $\Delta\tilde{\Gamma}(0) = \mathcal{G}(0) =
\Sigma^{\,r}_V(0) = 0$. Therefore
$\mF(0) = 1$ as it should.

Some 1PI diagrams
(Figs. \ref{fig.VFFl6}.a and \ref{fig.VFFl6}.e and the V terms in
Figs. \ref{fig.VFFl6}.b and \ref{fig.VFFl6}.c)
have a corresponding reducible counterpart involving
a vector propagator. The combination of both contributions can be then
incorporated in $A(q^2)$. The function $G(q^2,\mu^2)$
contains the corrections generated by the other 1PI diagrams
(Figs. \ref{fig.VFFl6}.d, \ref{fig.VFFl6}.f,
the S term in Fig.~\ref{fig.VFFl6}.b,
the S, A and P terms in Fig.~\ref{fig.VFFl6}.c
and the $\widetilde{\ell}_6$ and $\widetilde{r}^{\phantom{\, r}}_{V2}$ pieces in
Fig.~\ref{fig.VFFl6}.g).
Subtracting their contribution at $q^2=0$, which contains
the dependence on the renormalization scale $\mu$,
\bel{eq:G(0)}
G(0,\mu^2) \; =\;\Delta\Gamma (0,\mu^2)\; =\;
\Frac{1}{16\pi^2 F^2}\,\left\{
M_V^2\,\left[\Frac{3}{2}\,\ln{\Frac{M_V^2}{\mu^2}}+\Frac{1}{4}\right]
\, +\,
M_S^2\,\left[-\ln{\Frac{M_S^2}{\mu^2}}+\Frac{1}{2}\right]\right\}
\, ,
\ee
one gets:
\bear
\mathcal{G}(q^2) &=&
\Frac{C_0(q^2,M_V^2,M_V^2,0)}{F^2} \left[ \Frac{M_V^6}{q^2}+\Frac{M_V^4}{2}\right]
\; +\;
\Frac{C_0(q^2,M_S^2,M_S^2,0)}{F^2} \left[\Frac{M_S^6}{q^2}-\Frac{M_S^4}{2}\right]
\nn\\[10pt] &&
+\;\Frac{\overline{B}_0(q^2,M_V^2)}{F^2}
\left[-\Frac{M_V^4}{q^2}-\Frac{8M_V^2}{3}+\Frac{q^2}{4}+\Frac{q^4}{6M_V^2}\right]
\; +\;
\Frac{\overline{B}_0(q^2,M_S^2)}{F^2}
\left[-\Frac{M_S^4}{q^2}-\Frac{M_S^2}{3}+\Frac{q^2}{12}\right]
\nn\\[10pt] &&
+\;\Frac{\overline{B}_0(q^2,M_A^2)}{F^2}
\left[-2M_A^2-\Frac{q^2}{6}+\Frac{q^4}{6M_A^2}\right]
\; +\;
\Frac{\overline{B}_0(q^2,M_P^2)}{F^2}\left[-\Frac{2M_P^2}{3}+\Frac{q^2}{6}\right]
\\[10pt] &&
+\;\Frac{1}{16\pi^2 F^2}\;\left\{ \;\Frac{M_V^4 + M_S^4}{q^2}\, + \,\Frac{3}{4} M_V^2
\, -\, \Frac{1}{4} M_S^2
\right.\nn\\[10pt] &&\hskip 1.5cm
+\, q^2\,\left[\Frac{1}{12}\,\ln{\Frac{M_V^2}{\mu^2}} +
\Frac{1}{2}\,\ln{\Frac{M_A^2}{\mu^2}} -\Frac{1}{12}\,\ln{\Frac{M_S^2}{\mu^2}}
-\Frac{1}{6}\,\ln{\Frac{M_P^2}{\mu^2}} +\Frac{4}{9}
-16\pi^2\,\widetilde{\ell}_6^{r}(\mu)\right]
\nn\\[10pt] &&\hskip 1.5cm\left.
-\, \Frac{q^4}{6}\,\left[\Frac{1}{M_V^2}\,\ln{\Frac{M_V^2}{\mu^2}}
+\Frac{1}{M_A^2}\,\ln{\Frac{M_A^2}{\mu^2}}
+\Frac{4}{3} \left(\Frac{1}{M_V^2}+\Frac{1}{M_A^2}\right)
- \Frac{96\pi^2}{F^2}\,\widetilde{r}^{\, r}_{V2}(\mu)\right]
\;\right\}\, .
\nn\eear

\section{Low-Energy Limit}
\label{sec:LowE}

At very low energies, $q^2 \ll M_R^2$, the resonance fields can be integrated out
from the effective theory. One recovers then \cite{therole} the standard \chpt\
Lagrangian, which leads to the following result for the VFF \cite{MesonFF,op6FF}:
%
\bear\label{eq.VFFop4CHI}
\lefteqn{\mathcal{F}_{\chi\mathrm{PT}}(q^2) \, =\,
1\, -\,\Frac{q^2}{F^2}\,
\left\{\ell_6^{\,r}(\mu) \, +\,\Frac{1}{96\pi^2}\,
\left[\ln{\left( -\Frac{q^2}{\mu^2}\right)}-\Frac{5}{3}\right]
\right\}}&&
\\[10pt] &&\mbox{}
 +\,\Frac{q^4}{F^4}\, \left\{ r^{\, r}_{V2}(\mu)\,
+\,\Frac{1}{96\pi^2}\,
\left[\ln{\left( -\Frac{q^2}{\mu^2}\right)}-\Frac{5}{3}\right]\,
\left( 2 \ell_1^{\,r}-\ell_2^{\,r} + \ell_6^{\,r}\right)(\mu)\,
 + \,\cO\left(\Frac{1}{N_C^2}\right)\right\}
\, + \, \cO\left(\Frac{q^6}{F^6}\right) .
\nn\eear
The Taylor expansion in powers of $q^2$ of the \rcht\ prediction
\eqn{eq:VFF_result} reproduces
the \chpt\ formula \eqn{eq.VFFop4CHI}, as it should.
The coefficient of the $\cO\left[q^4\ln{(-q^2/\mu^2)}\right]$ term
satisfies the known large-$N_C$ equality \cite{therole,PI:02}\
$\left( 2 \ell_1-\ell_2 + \ell_6\right) \, =\, F^2 (1 - 5 M_S^2/M_V^2)/(2 M_S^2)$.
The non-logarithmic $\cO(q^4)$ and $\cO(q^6)$ terms relate the low-energy
chiral couplings $\ell_6$ and $r^{\phantom{\, r}}_{V2}$ with their \rcht\ counterparts
$\widetilde{\ell}_6$ and $\widetilde{r}^{\phantom{\, r}}_{V2}$:
\bear\label{eq:l6_rel}
\ell_6^{\,r}(\mu) & = & \mbox{}
- \Frac{F^2}{M_V^2}\, (1 +\hat{\delta}_V)\, +\,\widetilde{\ell}_6^{\,r}(\mu)
\, - \,\Frac{1}{96\pi^2}\,\left[
\ln{\Frac{M_V^2}{\mu^2}} - \ln{\Frac{M_P^2}{\mu^2}} + 3\,
\ln{\Frac{M_A^2}{\mu^2}} - \Frac{13}{6}\right]
\nn\\[10pt]
& = &\mbox{}
- \Frac{F_V G_V^{\,r}(\mu)}{M_V^2} \, +\,
\widetilde{\ell}_6^{\,r}(\mu)
\, +\,\Frac{1}{16\pi^2}\,\left[
\Frac{4}{3}\,\ln{\Frac{M_V^2}{\mu^2}}-
\Frac{1}{2}\,\ln{\Frac{M_A^2}{\mu^2}} +
\Frac{1}{6}\,\ln{\Frac{M_P^2}{\mu^2}} -
\Frac{M_S^2}{M_V^2}\,\ln{\Frac{M_S^2}{\mu^2}}
\right.\nn\\[10pt]&&\hskip 5.5cm\left.\mbox{}
+ \Frac{11}{18} + \Frac{M_S^2}{2 M_V^2}
\right]\, ,
\\[10pt]\label{eq:f1_rel}
r^{\, r}_{V2}(\mu) & = &
\Frac{F^2 F_V G_V^{\,r}(\mu)}{M_V^4}\, +\, \widetilde{r}^{\, r}_{V2}(\mu)
\, +\, \Frac{2 F^4}{M_V^2}\,\left[\hat{X}-X_Z^r(\mu)\right]
\nn\\[10pt]&&\mbox{}
+\,\Frac{F^2}{96\pi^2}\,\left\{
\left(6\,\Frac{M_S^2}{M_V^4}+\Frac{1}{2 M_V^2}
-\Frac{1}{2 M_S^2}\right)\,\ln{\Frac{M_S^2}{\mu^2}}
-\Frac{9}{M_V^2}\,\ln{\Frac{M_V^2}{\mu^2}}-
\Frac{1}{M_A^2}\,\ln{\Frac{M_A^2}{\mu^2}}
\right.\nn\\[10pt]&&\hskip 1.75cm\left.\mbox{}
 -\Frac{167}{60 M_V^2}- \Frac{17}{10 M_A^2}
- \Frac{3 M_S^2}{M_V^4} + \Frac{17}{20 M_S^2}
+ \Frac{1}{10 M_P^2}\right\}\, .
\eear
Notice that the combination of subleading \rcht\ couplings $\hat{X}$ does
not appear at $\cO(p^4)$. Therefore, the relation \eqn{eq:l6_rel} adopts
the same form in terms of the effective couplings defined in
\eqn{eq.effcouplings}, i.e.
$\widetilde{\ell}_6^{\,\mathrm{eff},r}(\mu) -
F_V^{\,\mathrm{eff}}\, G_V^{\,\mathrm{eff},r}(\mu)/(M_V^2)^{\mathrm{eff},r}(\mu) =
\widetilde{\ell}_6^{\,r}(\mu) -F_V\, G_V^{\,r}(\mu)/M_V^2$.
As shown in \eqn{eq:f1_rel}, this is no longer true at $\cO(p^6)$;
nevertheless, the explicit dependence on $\hat{X}-X_Z^r(\mu)$
present in $r^{\, r}_{V2}(\mu)$ can be reabsorbed into the leading term,
through the use of the effective couplings, i.e.
$r^{\, r}_{V2}(\mu) =
F^2 F_V^{\,\mathrm{eff}}\, G_V^{\,\mathrm{eff},r}(\mu)/(M_V^4)^{\mathrm{eff},r}(\mu)
+\widetilde{r}^{\, \mathrm{eff},r}_{V2}+\cdots$

Eqs.~\eqn{eq:l6_rel} and \eqn{eq:f1_rel} contain the well known lowest-order predictions
for the two \chpt\ couplings: $\ell_6 = -M_V^2 r^{\, r}_{V2}/F^2 = - F^2/M_V^2$.
Moreover, they give their dependence on the renormalization scale at the
NLO.  
The running of the renormalized couplings
[$\ell_6^{\,r}(\mu),r^{\, r}_{V2}(\mu)]$ and
[$\widetilde{\ell}_6^{\,r}(\mu),\widetilde{r}^{\, r}_{V2}(\mu)]$
is different, because their corresponding effective theories
have a very different particle content.

The $\mu$ dependence of a given coupling ``g'' can be characterized through the
logarithmic derivative
\bel{eq:gamma_F}
\mu\,\Frac{d g}{d\mu}\, =\, -\Frac{\gamma_g}{16\pi^2}\, .
\ee
From Eqs.~\eqn{eq:l6_run} and \eqn{eq:f1_run} one gets the running
of the \rcht\ couplings:
\bel{eq:gamma_rcht}
\gamma_{\strut\,\widetilde{\ell}_6}\, =\,\Frac{2}{3}
\qquad\quad , \qquad\quad
\gamma_{\strut\,\widetilde{r}^{\phantom{\, r}}_{V2}}\, =\,\Frac{F^2}{3}\,
\left(\Frac{1}{M_V^2}+\Frac{1}{M_A^2}\right)\, =\,
\Frac{F^2}{2 M_V^2}\, .
\ee
Eqs.~\eqn{eq:l6_rel} and \eqn{eq:f1_rel} give then the dependence
on the renormalization scale of
the corresponding \chpt\ couplings:
\bel{eq:gamma_chpt}
\gamma_{\strut\ell_6}\, =\,-\Frac{1}{3}
\qquad\quad , \qquad\quad
\gamma_{\strut r^{\phantom{\, r}}_{V2}}\, =\,\Frac{F^2}{6}\,
\left(\Frac{5}{M_V^2}-\Frac{1}{M_S^2}\right)\, .
\ee
These values are
in perfect agreement with the low-energy results of
refs.~\cite{chpt1loop,chptms,op6FF,op6lagrangian}.
The running of the $\cO(p^6)$ coupling $r^{\phantom{\, r}}_{V2}(\mu)/F^4$ receives of course
additional 2-loop contributions which are of $\cO(1/N_C^2)$.

The rigorous control of the renormalization scale dependences allows us to
investigate the successful resonance saturation approximation \cite{therole} at the NLO.
The \chpt\ couplings $\ell_6$ and $r^{\phantom{\, r}}_{V2}$ have been
phenomenologically extracted from a fit to the VFF data at low momenta.
This determines \cite{op6FF} the scale-invariant combination
\bel{eq:l6bar}
\bar \ell_6\,\equiv\, \Frac{32\pi^2}{\gamma_{\strut\ell_6}}\,\ell_6^{\,r}(\mu)
\, -\, \log{\Frac{m_\pi^2}{\mu^2}}\, =\,
16.0\pm 0.5\pm 0.7\, ,
\ee
and
\bel{eq:r2Mrho}
r^{\, r}_{V2}(M_\rho)\, = \, (1.6\pm 0.5)\cdot 10^{-4}\, .
\ee
Inserting these numbers in Eqs.~\eqn{eq:l6_rel} and \eqn{eq:f1_rel}, one can
estimate the corresponding scale-invariant combinations of NLO couplings
in \rcht:
\bel{eq:l6hat}
\hat{\ell}_6\,\equiv\,\widetilde{\ell}_6^{\,r}(\mu)
-\Frac{\gamma_{\strut\,\widetilde{\ell}_6}}{32\pi^2}\,\log{\Frac{M_V^2}{\mu^2}}
-\Frac{F^2}{M_V^2}\,\hat{\delta}_V\, ,
\ee
\bel{eq:rV2hat}
\hat{r}^{\phantom{\, r}}_{V2}\,\equiv\, \widetilde{r}^{\, r}_{V2}(\mu)
+ \Frac{F^4}{M_V^4}\,\left(\hat{\delta}_V + 2 M_V^2\,
\left[\hat{X}-X_Z^r(\mu)\right]\right)
- \Frac{\gamma_{\strut\,\widetilde{r}^{\phantom{\, r}}_{V2}}
-\Frac{2 F^4}{M_V^2}\,\gamma_{_{X_Z}}}{32\pi^2}\,
\log{\Frac{M_V^2}{\mu^2}}\, ,
\ee
where $\gamma_{_{X_Z}}=-1/(6 F^2)$. Taking $F=92.4$~MeV, $M_V = 770$~MeV
and $M_S = 1$~GeV, one gets
$\hat{\ell}_6 = (-0.2\pm 0.9)\cdot 10^{-3}$ and
$\hat{r}^{\phantom{\, r}}_{V2}= (-0.2\pm 0.5)\cdot 10^{-4}$,
while a larger value
of the scalar resonance mass $M_S = 1.4$~GeV shifts the $\cO(p^4)$
coupling to
$\hat{l}_6 = (-0.9\pm 0.9)\cdot 10^{-3}$, without affecting
$\hat{r}^{\phantom{\, r}}_{V2}$ at the quoted level of accuracy.
These numbers should be compared with the large--$N_C$ predictions
for the \chpt\ couplings \
$\ell_6|_{N_C\to\infty} = -F^2/M_V^2 = -0.014$ \ and \
$r^{\phantom{\, r}}_{V2}|_{N_C\to\infty} =F^4/M_V^4 = 2.1\cdot 10^{-4}$.
Put in a different way, the hypothesis
$\hat{\ell}_6 = \hat{r}^{\phantom{\, r}}_{V2} = 0$
generates excellent predictions for $\ell_6^{\,r}(\mu)$ and $r^{\, r}_{V2}(\mu)$
at any scale $\mu$.

\section{Behaviour at Large Energies}
\label{sec:LargeE}

At large momentum transfer, the relevant renormalization scale invariant
functions take the forms:
\bear
\mathcal{G}(q^2) &\!\! =&\!\!
\Frac{1}{16\pi^2 F^2}\,\left\{-\; q^4\,\left[
\Frac{1}{6}\,\left(\Frac{1}{M_V^2}+\Frac{1}{M_A^2}\right)\,
\left(\ln{\Frac{-q^2}{\mu^2}}-\Frac{2}{3}\right)
-\Frac{16\pi^2}{F^2}\,\widetilde{r}_{V2}^{\, r}(\mu)\right]
\right.\nn\\[10pt]&&\hskip 1.6cm\left.\mbox{}
+\, q^2\,\left[\Frac{1}{3}\,\ln{\Frac{-q^2}{\mu^2}}
+\Frac{16}{9} -16\pi^2\,\widetilde{\ell}_6^{\,r}(\mu)\right]\,
+\, \cO\left(1\right)\right\}\, ,
\eear
\be
\Delta\tilde{\Gamma}(q^2) \, =\,
\Frac{M_V^2}{16\pi^2 F^2}\,\left\{\ln{\Frac{-q^2}{M_V^2}}\,
\left[\ln{\Frac{q^2}{M_V^2}}-2\right]
-\Frac{1}{2}\,\ln^2{\Frac{q^2}{M_V^2}} -
\Frac{\pi^2}{6}+\Frac{9}{4}+\Frac{M_S^2}{4\, M_V^2}\right\}
+ \cO\left(\Frac{1}{q^2}\right) \, ,
\ee
\be
\Sigma^{\,r}_V(q^2)\, =\,  
\Frac{-q^4}{96\pi^2 F^2}\,\left\{
\ln{\Frac{-q^2}{\mu^2}}-\Frac{5}{3} + 192\pi^2 F^2\, X^r_Z(\mu)\right\}\, .
\ee

The $\rho$ propagator makes the $A(q^2)$ piece of the VFF well behaved when
$q^2\to\infty$. However, the 1PI contributions generate a wrong behaviour
$\mathcal{G}(q^2)\sim q^4\,\ln{(-q^2/\mu^2)}$ in the $B(q^2)$ term, which
cannot be eliminated with a local contribution.
The problem originates in the two-resonance cut which has an unphysical
growing with momenta.
Although our leading \rcht\ Lagrangian \eqn{eq.LRint} only incorporates 
couplings linear in the resonance fields,
the kinetic resonance Lagrangian \eqn{eq.LRkin} introduces some bilinear
interactions through the chiral connection included in the covariant
derivatives. Their couplings are fixed by chiral symmetry and give rise
to the 1PI diagrams in Figs.~\ref{fig.VFFl6}.c, \ref{fig.VFFl6}.d
and \ref{fig.VFFl6}.f.
Obviously, these are not the only interactions bilinear in the resonance
fields even at large--$N_C$ \cite{RPP:03,CEEPP:04,GPP:04,CEEPPK:04}.
Therefore, it is not surprising that
our calculation is unable to find the correct behaviour at large energies
for those contributions with two intermediate resonances.

The contributions with an internal vector propagator in diagrams
\ref{fig.VFFl6}.b and \ref{fig.VFFl6}.c give us some hint about
which pieces could be missing in our calculation. These two diagrams
combine with a reducible contribution of the type \ref{fig:topologies}.b:
the 1PI $\langle V_{\mu\nu}\pi\pi\rangle$ vertex in Fig.~\ref{fig.GV}.b.
The three contributions contain identical loop functions and their sum
generates a global factor $M_V^2/(M_V^2 - q^2)$, which suppresses
the large--$q^2$ behaviour. Thus, these corrections have been included
in the term $A(q^2)$.

It seems natural to conjecture that the remaining 1PI contributions
with two-resonance cuts should combine with the corresponding
reducible topologies, including $\langle V RR\rangle$
and $\langle v^\mu RR\rangle$ vertices,
to generate the final propagator suppression:
\bel{eq:conjecture}
G(q^2) \quad\longrightarrow\quad
\Frac{M_V^2}{M_V^2 - q^2 -\Sigma^{\,r}_V(q^2)}\;\; G(q^2)\, .
\ee
The needed Lagrangian takes the form
\be
\mL^2_{VRR}
\, =  \, i \, \lambda^{VSS}\,\bra V^{\mu\nu}\,\nabla_\mu S\,\nabla_\nu
S \ket\; +\; i \, \lambda^{VPP}\,\bra V^{\mu\nu}\,\nabla_\mu P\,
\nabla_\nu P\ket\; +\;\cdots
\ee
Our conjecture fixes the new chiral couplings in the large--$N_C$ limit.
It would be interesting to analyze the contributions of this Lagrangian
to appropriate Green functions, following the work of
refs.~\cite{KN:01,RPP:03,CEEPP:04}, and check whether the couplings
predicted by the corresponding short-distance QCD corrections
agree with our naive conjecture.
In appendix C, we show two simple examples where the presence of the
propagator suppression can be demonstrated in a rather straightforward way.

The behaviour at large energies is also constrained by unitarity requirements.
Moreover, the local contributions can be forced to vanish at large $q^2$ by taking
appropriate values of the \rcht\ couplings. Probably, this could allow us to
determine the scale invariant constants $\hat{\ell}_6$ and
$\hat{r}^{\phantom{\, r}}_{V2}$.
We plan to investigate all these points in forthcoming works.

\section{Summary}
\label{sec:Summary}

The one-loop analysis of the VFF has shown a series of interesting features.
As expected, loop diagrams with massive resonance states in the internal lines
generate ultraviolet divergences, which require additional higher-dimensional
counterterms in the \rcht\ Lagrangian. Since these counterterms give rise to
tree-level contributions which grow too fast at large momenta, their corresponding
couplings should be zero at leading order in the large--$N_C$ expansion.
Thus, one can establish a well defined counting in powers of $1/N_C$ to organize
the calculation.

The formal renormalization is completely straightforward at one loop.
One can easily determine the $\mu$ dependence of all relevant renormalized
couplings. Moreover, the final result is only sensitive to some combinations
of the chiral couplings. In fact, using the lowest-order equations of motion,
one can eliminate most of the higher-order couplings. Their effects get then
reabsorbed into redefinitions of the lowest-order parameters.

Expanding the result in powers of $q^2/M_R^2$, one recovers the usual \chpt\
expression at low momenta. This relates the low-energy chiral couplings
$\ell_6$ and $r^{\phantom{\, r}}_{V2}$ with their corresponding \rcht\ counterparts
$\widetilde{\ell}_6$ and $\widetilde{r}^{\phantom{\, r}}_{V2}$.
The rigorous control of the
renormalization scale dependences has allowed us to investigate the successful
resonance saturation approximation at the next-to-leading order in $1/N_C$.
The assumption $\hat{\ell}_6 = \hat{r}^{\phantom{\, r}}_{V2} = 0$
generates excellent predictions for $\ell_6^{\,r}(\mu)$ and $r^{\, r}_{V2}(\mu)$
at any scale $\mu$.

At high energies, we have identified a problematic behaviour which originates
in the two-resonance cuts: they generate an unphysical increase of the VFF at
large values of momentun transfer.
This is not surprising, since there are additional contributions generated
by interaction terms with several resonances, which have not been included in
the minimal \rcht\ Lagrangian.
These new chiral structures should be taken into account to achieve a 
physical description of the VFF above the two-resonance thresholds.
The short-distance QCD constraints can be used
to determine their corresponding couplings.

Our calculation represents a first step towards a systematic procedure to
evaluate next-to-leading order contributions in the $1/N_C$ counting.
More work in this direction is in progress.

\section{Acknowledgements}
We have benefited from useful discussions with Oscar Cat\`a, Gerhard Ecker,
Santi Peris, Ximo Prades and Jorge Portol\'es.
This work has been supported in part by
the EU HPRN-CT2002-00311 (EURIDICE), by MCYT (Spain) under grant
FPA-2001-3031 and by ERDF funds from the EU Commission.

\newpage

\appendix
\newcounter{ap1}
\renewcommand{\thesection}{\Alph{ap1}}
\renewcommand{\theequation}{\Alph{ap1}.\arabic{equation}}
\setcounter{ap1}{1}
\setcounter{equation}{0}
\hspace*{-0.6cm}{\large \bf Appendix A: Feynman Integrals}

The calculation involves the following Feynman Integrals:
\be
A_0(M^2)\,\equiv\,\int \Frac{dk^d}{i(2\pi)^d}\;
\Frac{1}{k^2+i\epsilon \, - \, M^2}
\; =\;
- \,\Frac{M^2}{16\pi^2}\,\left[ \lambda_\infty \,
+ \, \ln{\Frac{M^2}{\mu^2}}\right] \, ,
\ee
\bear
\lefteqn{B_0(q^2,M_a^2,M_b^2) \,\equiv\,\int \Frac{dk^d}{i(2\pi)^d}\;
\Frac{1}{(k^2+i\epsilon \, - \, M_a^2) \,
[(q-k)^2+i\epsilon \, - \, M_b^2]}} &&
\\[10pt] &&\qquad =
- \,\Frac{1}{16\pi^2}\,\left[ \lambda_\infty \, + \,
\Frac{M_a^2}{M_a^2 - M_b^2}\ln{\Frac{M_a^2}{\mu^2}}\, - \,
\Frac{M_b^2}{M_a^2 - M_b^2}\ln{\Frac{M_b^2}{\mu^2}}\right] \, +\,
\bar{J}(q^2,M_a^2,M_b^2)
\, ,
\nn\eear
and the finite function
\bear
\lefteqn{C_0(q^2,M_a^2,M_b^2,M_c^2) \,\equiv} &&
\\[10pt] &&\qquad =
\int \Frac{dk^d}{i(2\pi)^d}\;
\Frac{1}{[(p_1-k)^2+i\epsilon \, - \, M_a^2] \,
[(p_2+k)^2+i\epsilon \, - \, M_b^2] \,
(k^2+i\epsilon \, - \, M_c^2)} \; ,
\nn\eear
with $q=p_1+p_2$ and, with massless outgoing pions, $p_1^2=p_2^2=0$.
The divergences are collected in the factor
\be
\lambda_\infty\, =\,   \Frac{2 \,\mu^{d-4}}{d-4}\, +\, \gamma_E\,
-\, \ln 4\pi\, - 1\, ,
\ee
being \ $\gamma_E\simeq 0.5772\ldots$ \ the Euler constant and $\mu$ the renormalization scale.

The two-propagator integral contains the finite function
\bear
\bar{J} (q^2,M_a^2,M_b^2) & = &
\frac{1}{32 \pi^2} \left\{ 2 + \left[
\frac{M_a^2-M_b^2}{q^2} - \frac{M_a^2+M_b^2}{M_a^2-M_b^2}\right]\,
\ln \frac{M_b^2}{M_a^2}
\right.\\[10pt]&&\left.
-\,\frac{\lambda^{1/2} (q^2,M_a^2,M_b^2)}{q^2} \;
\ln{\left( \frac{[q^2 + \lambda^{1/2} (q^2,M_a^2,M_b^2)]^2 -
(M_a^2-M_b^2)^2}{[q^2 -\lambda^{1/2} (q^2,M_a^2,M_b^2)]^2 -
(M_a^2-M_b^2)^2} \right) } \right\}\, ,
\nn\eear
with \
$\lambda (x,y,z)\,\equiv\, x^2 + y^2 + z^2 - 2 x y - 2 x z - 2 y z$\ .
Some useful particular cases are:
\bear
B_0(q^2,0,0)& = & - \,\Frac{\lambda_\infty}{16\pi^2}
\, + \, \hat{B}_0(q^2/\mu^2)\, ,
\nn\\[10pt]
B_0(q^2,M^2,M^2)& = & -\, \Frac{1}{16\pi^2}\left\{\lambda_\infty
+\ln{\Frac{M^2}{\mu^2}} + 1 \right\}
\, + \, \overline{B}_0(q^2,M^2) \, ,
\\[10pt]
B_0(q^2,0,M^2)& = & -\, \Frac{1}{16\pi^2}\left\{\lambda_\infty
+\ln{\Frac{M^2}{\mu^2}} \right\}
\, + \, \bar{J}(q^2,0,M^2) \, ,
\nn\eear
with the finite parts
\bear
\hat{B}_0(q^2/\mu^2)& = & \Frac{1}{16\pi^2}\,\left\{1
-\ln{\left(-\Frac{q^2}{\mu^2}\right)}  \right\} \, ,
\nn\\[10pt]
\overline{B}_0(q^2,M^2)& \equiv &
\bar{J}(q^2,M^2,M^2) \, =\,
\Frac{1}{16\pi^2}\left\{2 - \sigma_M
\ln{\left(\Frac{\sigma_M+1}{\sigma_M-1}\right)} \right\} \, ,
\\[10pt]
\bar{J}(q^2,0,M^2)& \equiv &
\Frac{1}{16\pi^2}\left\{ 1 -\left( 1 -{M^2\over q^2}\right)\,
\ln{\left(1 -{q^2\over M^2}\right)}\right\}\, ,
\nn\eear
where \ $\sigma_M=\sqrt{1-4 M^2/q^2}$.

The relevant three-propagator integrals are:
\bear
C_0(q^2,0,0,M^2) &=& -\,\Frac{1}{16 \pi^2 q^2}\,
\left\{ \mbox{Li}_2\left(1+\Frac{q^2}{M^2}\right) -\, \mbox{Li}_2(1)\right\} \, ,
\nn\\[10pt]
C_0(q^2,M^2,M^2,0) &=&
\Frac{1}{16 \pi^2 q^2}\;\ln^2{\left({\sigma_M-1\over\sigma_M+1}\right)}\, ,
\eear
where
\be
\mbox{Li}_2(y)\,\equiv\,
- \,  \int_0^1\, \Frac{dx}{x}\; \ln{ (1-xy) }
\, = \, - \, \Int_0^y\, \Frac{dx}{x}\; \ln{ (1-x) }
\ee
is the usual dilogarithmic function.
%
%

\appendix
\newcounter{ap2}
\renewcommand{\thesection}{\Alph{ap2}}
\renewcommand{\theequation}{\Alph{ap2}.\arabic{equation}}
\setcounter{ap2}{2}
\setcounter{equation}{0}
\vspace*{0.5cm}
\hspace*{-0.6cm}{\large \bf Appendix B: Lorentz Structures in the Vector
Propagators}

In momentum space, the bare vector-field propagator can be written in the form
\be
\bra V^{\mu\nu}V^{\rho\sigma}\ket_0 \, = \,
\Delta^{\mu\nu,\rho\sigma}(q) \, = \, \Frac{2i}{M_V^2-q^2} \;
\mA^{\mu\nu,\rho\sigma}(q)
\, + \, \Frac{2i}{M_V^2} \; \Omega^{\mu\nu,\rho\sigma}(q) \, ,
\ee
with the antisymmetric tensors
\bear
\mA_{\mu\nu,\rho\sigma}(q) &\equiv &\Frac{1}{2q^2} \,
\left[\, g_{\mu\rho}q_\nu q_\sigma -
g_{\rho\nu}q_\mu q_\sigma - (\rho \leftrightarrow \sigma) \,\right]\, ,
\nn\\[10pt]
\Omega_{\mu\nu,\rho\sigma}(q) &\equiv & -\,\Frac{1}{2q^2}\,
\left[ g_{\mu\rho} q_\nu q_\sigma -
g_{\rho\nu}q_\mu q_\sigma  - q^2 g_{\mu\rho}g_{\nu\sigma} -
(\rho\leftrightarrow\sigma) \,\right]\, ,
\\[10pt]
\mI_{\mu\nu,\rho\sigma} &\equiv & \Frac{1}{2}\, \left(
g_{\mu\rho} g_{\nu\sigma} - g_{\mu\sigma} g_{\nu\rho}\right) \, ,
\nn\eear
obeying the properties:
\be
\ba{c}
\Omega\cdot \mA\, =\, \mA\cdot \Omega\, =\, 0 \quad\; , \quad\;
\mA\cdot \mA\, = \,\mA \quad\; , \quad\;
\Omega \cdot \Omega\, = \,\Omega \quad\; , \quad\;
\mA \,+\,\Omega\, = \,\mI \, ,
\\[10pt]
q^\mu\,\Omega_{\mu\nu,\rho\sigma}(q) \; = \;
q^\nu\,\Omega_{\mu\nu,\rho\sigma}(q) \; = \;
q^\rho\,\Omega_{\mu\nu,\rho\sigma}(q) \; = \;
q^\sigma\,\Omega_{\mu\nu,\rho\sigma}(q) \; = \; 0\, .
\ea
\ee

For any antisymmetric tensor $H_{\mu\nu,ab}$, the operator
$\mI_{cd,\alpha\beta}$ acts like the identity, i.e.
\be
H\cdot \mI \, =\, \mI \cdot H \, =\, H \, .
\ee
We can then define the antisymmetric inverse $G^{ab,\rho\sigma}$,
which satisfies
\be
H_{\mu\nu,ab}\; G^{ab,\rho\sigma} \,=\,
G_{\mu\nu,ab}\; H^{ab,\rho\sigma} \, = \, \mI_{\mu\nu}^{\rho\sigma}\, .
\ee
The inverse propagator in momentum space is given by
\bear
{\Delta^{-1}(q)}^{\mu\nu,\rho\sigma} &  = &
-\, i\,\Frac{\left(M_V^2-q^2\right)}{2}\;
 \mA^{\mu\nu,\rho\sigma}(q)
\, - \, i\,\Frac{M_V^2}{2}\; \Omega^{\mu\nu,\rho\sigma}(q)
\nn\\[10pt] &=&
i\,\Frac{q^2}{2} \; \mA^{\mu\nu,\rho\sigma}(q) \,
- \, i\,\Frac{M_V^2}{2} \;\mI^{\mu\nu,\rho\sigma} \, .
\eear
%

\appendix
\newcounter{ap3}
\renewcommand{\thesection}{\Alph{ap3}}
\renewcommand{\theequation}{\Alph{ap3}.\arabic{equation}}
\setcounter{ap3}{3}
\setcounter{equation}{0}
\vspace*{0.5cm}
\hspace*{-0.6cm}{\large \bf Appendix C: Form Factors with Resonances in the
Final State}

We present here a few examples of current matrix elements with external resonances.
They show that in order to implement a correct short-distance behaviour, one needs
to introduce additional interactions with more than one resonance field. Moreover,
the new chiral couplings can be easily determined.

\vspace*{0.5cm}
\noindent {\bf C.1 \ Axial form factor to  $\mathbf{S^0_{I=0}\,\pi^-}$}

\vspace*{0.2cm}

Let us consider the two-point correlation function of two axial currents
$J_A^\mu = \bar{d}\gamma^\mu \gamma_5 u$,
in the chiral limit:
\bel{eq:Acorr}
\Pi_{AA}^{\mu\nu}(q)\,\equiv\, i\int d^4x \;\,
\mathrm{e}^{iqx}\;\langle 0|T\left(J_A^\mu(x)
J_A^\nu(0)^\dagger\right)|0\rangle
\, =\, \left( -g^{\mu\nu} q^2 + q^\mu q^\nu\right)\,\Pi_{AA}(q^2)\, .
\ee
The associated spectral function Im $\Pi_{AA}(t)$ is a sum of positive
contributions corresponding to the different intermediate states. At
large $t$, it behaves as a constant. Therefore, since there is an infinite
number of possible states,
the absorptive contribution of a given intermediate state should vanish at infinite
momentum transfer.

\begin{figure}[t]\centering
\includegraphics[height=12cm,angle=-90,clip]{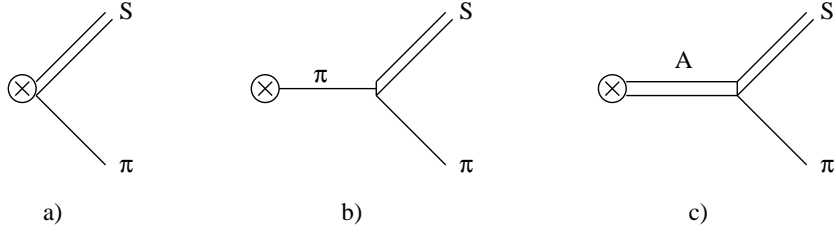}
\caption{Tree-level contributions to $\mF_{S\pi}(q^2)$.}
\label{fig:FSpi}
\end{figure}

One can easily check that the minimal \rcht\ Lagrangian of ref.~\cite{therole},
which only contains interactions linear in the resonance fields,
generates an absorptive $S^0_{I=0}\pi^-$ contribution
with the wrong behaviour at large momenta:
$\mbox{\rm Im}\Pi_{AA}(t)|_{S\pi}\,\sim\, $ constant.
The problem can be easily identified analysing the corresponding
form factor, defined through the matrix element
\bel{eq:FFSp}
\bra S^0_{I=0}\, \pi^-|\bar{d}\gamma_\mu \gamma_5 u| 0 \ket \, =
\,-  \, 2\, i \, \mF_{S\pi}(q^2) \;
\left(g_{\mu\nu}-\Frac{q_\mu q_\nu}{q^2}\right)\,
p_{\pi}^\nu\, ,
\ee
where $q^\mu= (p_{\pi}+p_{S})^\mu$.
The lowest-order calculation with the \rcht\ Lagrangian
(diagrams \ref{fig:FSpi}.a and  \ref{fig:FSpi}.b)
gives a constant form factor,
\be
\mF_{S\pi}(q^2)\, = \, \Frac{2 \,  c_d}{F} \, ,
\ee
which obviously is not vanishing at infinite momentum transfer.

The correct large energy behaviour can be recovered adding the
interaction term \cite{CEEPPK:04}
\be
\mL^2_{SA}\, = \, \lambda_1^{SA}\,\bra \left\{\nabla^\mu S,
A_{\mu\nu}\right\}  u^\nu\}\ket
\, ,
\ee
which modifies the form factor (diagram \ref{fig:FSpi}.c):
\be
\mF_{S\pi}(q^2)\, = \, \Frac{2c_d}{F} \,
- \, \sqrt{2}  \,\lambda_1^{SA}\, \Frac{F_A}{F}  \, \Frac{q^2}{M_A^2-q^2}\, .
\ee
Imposing that the form factor must vanish as $q^2\to \infty$,
the coupling $\lambda_1^{SA}$ is constrained to take the value
\be
\lambda_1^{SA}\, = \, - \, \Frac{\sqrt{2}\, c_d}{F_A}
\, =\,-\,\Frac{1}{\sqrt{2}}\, .
\ee
The resulting form factor adopts then the usual
monopolar form
\be
\mF_{S\pi}(q^2)\, = \, \Frac{M_A^2}{M_A^2-q^2}\, .
\ee
%

\vspace*{0.5cm}
\noindent
{\bf C.2 \ Vector form factor to $\mathbf{R^0_{I=1}\, R^-}$ \
($\mathbf{R=S,\, P}$)}

\vspace*{0.2cm}

\begin{figure}[t]\centering
\includegraphics[height=8cm,angle=-90,clip]{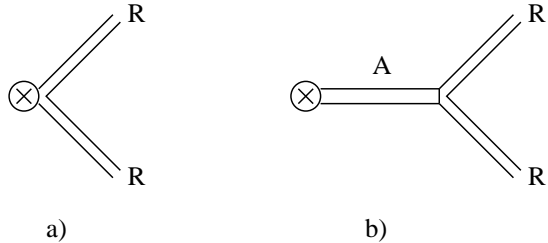}
\caption{Tree-level contributions to $\mF_{RR}(q^2)$.}
\label{fig:FRR}
\end{figure}

The two-point correlation function of two vector currents
has a similar behaviour at short distances. Its
spectral function behaves as a constant at large momentum transfer,
implying that the form factors associated with each intermediate
state should vanish at $q^2\to \infty$.

The minimal \rcht\ Lagrangian generates the matrix elements:
\be
\bra R^0_{I=1}(p_ 1)\, R^-(p_2)|\,\bar{d}\gamma^\mu u\, |0\ket \, =
\, \sqrt{2} \, (p_2 - p_1)^\mu \, \mF_{RR}(q^2) \, ,
\ee
where $R=S,P$ stands for a scalar or a pseudo-scalar resonance. The
corresponding form factors are just constant at lowest order
(diagram \ref{fig:FRR}.a):
\be
\mF_{RR}(q^2) \, = \, 1 \, .
\ee
This constant contribution originates in the chiral connection of the 
resonance kinetic Lagrangians.

It is possible again to recover the right QCD short distance
behaviour by adding the interaction terms
\be
\mL^2_{VRR}
\, =  \, i \, \lambda^{VRR}\,\bra V^{\mu\nu}\,\nabla_\mu R\,\nabla_\nu
R \ket \, ,
\ee
which change the form factors to (diagram \ref{fig:FRR}.b)
\be
\mF_{RR}(q^2) \, = \, 1 \, +\, \Frac{F_V}{\sqrt{2}}\, \lambda^{VRR}\,
\Frac{q^2}{M_V^2-q^2}  \, .
\ee
Imposing a proper high energy behaviour one gets the constraint
\be
\lambda^{VRR}\, =\, \Frac{\sqrt{2}}{F_V} \, =\,\Frac{1}{F} \, ,
\ee
and a monopolar form for the form factors
\be
\mF_{RR}(q^2)\, = \, \Frac{M_V^2}{M_V^2-q^2}\, .
\ee
%


\end{document}